\documentclass[a4paper,10pt,twocolumns]{article}
\pdfoutput=1

\usepackage{graphicx}
\usepackage{grffile}% allow spaces in graphicx filenames
\usepackage{xcolor}
\usepackage{url}
\usepackage[pdftex,colorlinks=true,linkcolor=blue,citecolor=blue,urlcolor=blue]{hyperref}% urlcolor=blue (default: magenta)
\usepackage[expproduct=cdot]{siunitx} %expproduct=tighttimes
\usepackage{enumitem} %\begin{enumerate}[itemsep=2pt,parsep=2pt,topsep=0pt,partopsep=0pt]
\usepackage{epstopdf}

\usepackage{relsize} % for relative font sizes, useful in commands such as PCa, Wnt\
\usepackage{amsmath}
\usepackage{amssymb}
\usepackage{bbm} % for \mathbbm{1}
\usepackage[utopia,greekuppercase=italicized]{mathdesign}\edef\partial{\mathchar\number\partial\noexpand\!} % variant in case newtxtext is not available (eg texlive 2011/arxiv). Loads adjuted times font through ptm.

\usepackage[T1]{fontenc} % better to use type 1 fonts than bitmapped fonts

\usepackage{ctable} % \toprule, \thickmidrule, \midrule, \cmidrule and \bottomrule to replace \hline in tables
\usepackage{tabularx} % For extensible columns. Alternative: tabulary. Calls \usepackage{array}
\usepackage[numbers,compress]{natbib}
\definecolor{royalblue4}{HTML}{27408B}% royalblue4 in Emacs (see M-x list-colors-display)
\definecolor{red4}{HTML}{8B0000}% red4 in Emacs
\definecolor{green4}{HTML}{008b00} % green4 in Emacs

\usepackage{dblfloatfix} % to use modifiers [h] etc. in \begin{figure*} ... \end{figure*} environment
\usepackage{fixltx2e} % prevent starred-figures from appearing out of ascending order
\usepackage[font=footnotesize,labelfont=bf,labelsep=endash,margin=0pt]{caption} % have figure/table caption in smaller font with bf marker
\usepackage[title,header]{appendix}

\newlength{\myleftmargin} 
\newlength{\mytopmargin} 
\newlength{\myrightmargin} 
\newlength{\mybottommargin} 
\usepackage[runin]{abstract}

\let\paragraphold\paragraph
\renewcommand*{\paragraph}[1]{\paragraphold{#1.}} % automatically add a dot after argument of \paragraph
\newcommand{\keywords}[1]{\vspace{2mm}\noindent\textbf{Keywords:} #1} % best to use within abstract
\newcommand{\pagewidetitle}[3] % to display #1=\maketitle, #2=abstract, and #3=title/author footnotes at the correct place
{%
    \twocolumn%
        [%
            \vskip-5mm%
            \begin{@twocolumnfalse}%
                #1%
                #2%
                \vspace{5mm}%
            \end{@twocolumnfalse}%
        ]%
        #3%
}

\newlength{\figurewidth}

\newcommand{\etal}{{\it et\ al.}}

\newcommand{\der}{\mathrm{d}}

\newcommand{\pd}[2]{\frac{\partial #1}{\partial #2}}

\newcommand{\td}[2]{\frac{\der #1}{\der #2}}

\newcommand{\Order}{\mathrm{O}}

\newcommand{\e}{\mathrm{e}}
\renewcommand{\b}[1]{{\boldsymbol{#1}}} % usually italic bold (greek letter etc.)

\newcommand{\da}{\ensuremath{\text{day}}} %std?
\newcommand{\days}{\ensuremath{\text{days}}} %std?
\newcommand{\um}{\ensuremath{\micro\metre}}%  micro meter, from siunitx
\usepackage{relsize} % for relative font sizes, useful in commands such as PCa, Wnt\
\newcommand{\kform}{\text{$k_\text{f}$}} % alternative to \ensuremath

\newcommand{\dperp}[2]{\left.\frac{\der #1}{\der #2}\right)_\perp}

\begin{document}
	\title{\bf Modelling the effect of curvature on the collective behaviour of cells growing new tissue}
	\renewcommand{\thefootnote}{\fnsymbol{footnote}}%
	\author{Almie Alias$^\text{a,b,}$\footnotemark[1], Pascal R Buenzli$^\text{a}$}
	
	\date{\normalsize \vspace{-2mm}$^\text{a}$School of Mathematical Sciences, Monash University, Clayton VIC 3800, Australia\\$^\text{b}$School of Mathematical Sciences, National University of Malaysia, 43600 Bangi, Selangor D. Ehsan, Malaysia\\\vskip 1mm \normalsize	
	\today\vspace*{-5mm}}
	
	\pagewidetitle{% remove this for standard LaTeX, or add \newcommand{\pagewidetitle}[1]{#1}
		\maketitle
	}{
	\begin{abstract}
The growth of several biological tissues is known to be controlled in part by local geometrical features, such as the curvature of the tissue interface. This control leads to changes in tissue shape that in turn can affect the tissue's evolution. Understanding the cellular basis of this control is highly significant for bioscaffold tissue engineering, the evolution of bone microarchitecture, wound healing, and tumour growth. While previous models have proposed geometrical relationships between tissue growth and curvature, the role of cell density and cell vigor remains poorly understood. We propose a cell-based mathematical model of tissue growth to investigate the systematic influence of curvature on the collective crowding or spreading of tissue-synthesising cells induced by changes in local tissue surface area during the motion of the interface. Depending on the strength of diffusive damping, the model exhibits complex growth patterns such as undulating motion, efficient smoothing of irregularities, and the generation of cusps. We compare this model with in-vitro experiments of tissue deposition in bioscaffolds of different geometries. By accounting for the depletion of active cells, the model is able to capture both smoothing of initial substrate geometry and tissue deposition slowdown as observed experimentally.

	\end{abstract}
}{
\protect\footnotetext[1]{Corresponding author. Email address: \texttt{almie.alias@monash.edu}}%
\renewcommand{\thefootnote}{\arabic{footnote}}%
}% end of \pagewidetitle. Remove this for standard LaTeX

%\maketitle 
%\tableofcontents

\section{Introduction}
Substrate geometry is an influential variable for new tissue growth with high significance for bioscaffold tissue engineering~\cite{Ripamonti2010}. Surface curvature~\cite{Dunn1976,Curtis1964}, and roughness~\cite{Martin1995,Deligianni2000} have important effects on cell behaviour in addition to the surface's chemical composition~\cite{Boyan1996,Gaudet2003,Cavalcanti-Adam2007,Arnold2009} and rigidity \cite{Lo2000,Pelham1997}. % Cells cultured within a three-dimensional matrix behave differently from cells cultured on a surface~\cite{Benton2009}.
At a single cell scale, tissue geometry affects the formation of focal adhesions on the cell membrane, resulting in differences in cell orientation, motility, shape, phenotype, and apoptosis due to biochemical and mechanical  effects~\cite{Clark1990,Curtis1998,Vogel2006,Parker2002,Thery2006,Chen1997,Yim2010}.

Larger geometrical features of substrates, that span multiple cell sizes, also influence tissue growth because they affect the collective behaviour of cell populations. Direct and indirect (e.g.\ mechanics-mediated) effects of tissue geometry on tissue growth are expected to play an important role in bone, tissue engineering, wound healing~\cite{Rolli2012,Poujade2007} and in tumour growth~\cite{Lowengrub2010}. Neotissue secreted by pre-osteoblasts cultured on porous scaffolds of various shapes grows at a rate that correlates with the local mean curvature~\cite{Nelson2005,Rumpler2008,Dunlop2010,Bidan2012,Bidan2013,Bidan2013b,Gamsjager2013,Knychala2013,Guyot2014,Guyot2015}. Such mean curvature flow leads to smoothing of the initial substrate geometry~\cite{Grayson1987, Crandall1983}. New bone deposition \emph{in vivo} occurs at different rates in compact cortical bone and porous trabecular bone, suggested to be due to the different substrate geometries in these tissues~\cite{Martin2000}. In contrast to \emph{in-vitro} tissue growth, cylindrical cavities in cortical bone infill at rates that correlate with the inverse mean curvature, i.e., tissue deposition slows down as infilling proceeds~\cite{Lee1964,Manson1965,Frost1969,Metz2003}. At the same time, irregularities of the initial substrate smooth out with tissue deposition: Haversian canals are more regular than osteon boundaries~\cite{Parfitt1983}.

These conflicting observations on the role of geometry in tissue growth may be reconciled if one takes into consideration the cellular basis of new tissue deposition, in particular cell density and cell vigor (new tissue synthesis rate)~\cite{Buenzli2014a}, and the various biological and geometrical influences that these variables are subjected to. A decrease in active cell number, due for example to quiescence, cell death, or detachment from the tissue surface~\cite{Jilka1999}, could explain tissue deposition slowdown. At the same time, local inhomogeneities in cell density and in cell vigor could explain smoothing of corners and irregularities. 

Previous mathematical models of the evolution of the tissue interface have proposed to capture the smoothing dynamics of \emph{in-vitro} tissue growth through a simple mathematical relation between interface velocity and mean curvature by comparing cell tension with surface tension problems in physics~\cite{Rumpler2008,Bidan2012,Bidan2013,Bidan2013b,Guyot2014,Guyot2015}. However, these geometric models do not account for cell numbers, which limits the interpretation of underlying biological processes. Part of the tissue growth slowdown observed \emph{in vitro} in 2D cross-sections has been tentatively explained by scaffold boundary effects leading to a catenoid tissue surface of smaller mean curvature than a cylindrical surface~\cite{Bidan2013,Bidan2013b}. The influence of cellular processes (such as a reduction in active cells or in cell vigor) cannot be factored in easily into these geometric models. In cortical bone formation \emph{in-vivo}, tissue surface is mostly cylindrical or conical and has moving  boundaries~\cite{Pazzaglia2010,Martin1998}. A slowdown of tissue deposition due to cellular processes rather than three-dimensional geometrical effects is more likely. Both surface cell density and cell vigor decrease during cortical infilling~\cite{Parfitt1994,Marotti1976,Buenzli2014a}.

In this paper, we develop a mathematical model of the effect of local curvature on the collective behaviour of cells synthesising new tissue at the tissue interface. We compare numerical simulations of the model with tissue growth dynamics in bioscaffolds of different pores shapes obtained in Refs~\cite{Bidan2012,Bidan2013}. This comparison suggests that a reduction in the number of active cells is a likely explanation for tissue deposition slowdown observed in these experiments.

The main purpose of the mathematical model is to determine the systematic influence of curvature on cell density due to the contraction or expansion of the local surface area during the evolution of the tissue interface. This influence is an inevitable geometrical pull: the deposition of new tissue on concave regions of the substrate reduces the local surface area, and so tends to increase surface cell density and crowd tissue; the deposition of new tissue on convex regions of the substrate increases the local surface area, and so tends to decrease surface cell density and spread tissue (Fig.~\ref{figCell}a). This systematic influence of curvature is important to elucidate and to single out, so that other geometrical influences on tissue growth can be determined, such as influences on individual cell vigor.

\section{Mathematical model}
\label{Sect2}
\begin{figure*}[h!]
% \captionsetup{justification=centering}

\centerline{
\includegraphics[width=\linewidth,clip]{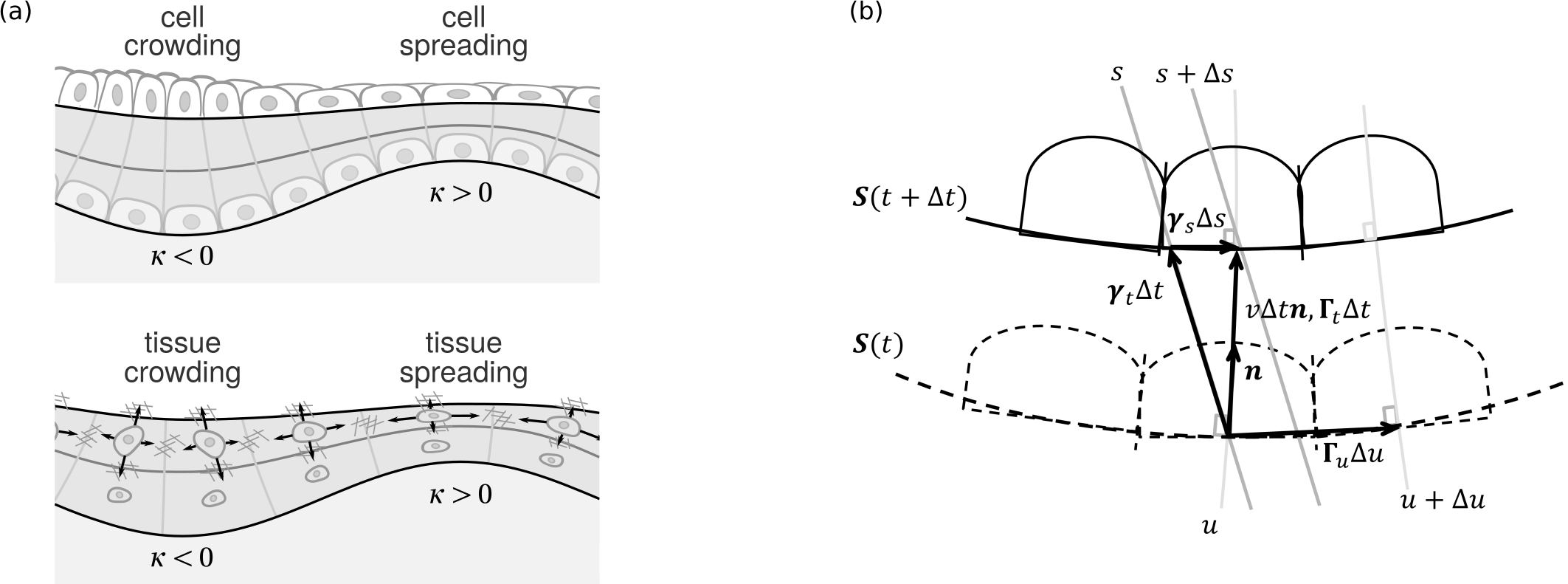}
}

\caption{(a) Cells lining a tissue surface such as bone-forming osteoblasts will concentrate or spread during the evolution of the interface depending on whether the initial substrate is concave ($\kappa<0$) or convex ($\kappa > 0$) (top). In a similar way, cellular and extracellular tissue volume produced near the tissue surface will crowd or spread depending on the substrate curvature (bottom). In both cases, this influences the local tissue growth rate. (b) Schematic diagram depicting the representation of the tissue surface $S(t)$ by an arbitrary parameterisation $\b\gamma(s,t)$ and by an orthogonal parameterisation $\b\Gamma(u,t)$. Timelines of $\b\Gamma(u,t)$ follow the cell's trajectories assumed normal to $S(t)$ at all times.}
\label{figCell}
\end{figure*}

We consider a biological tissue that grows by deposition of new matrix secreted by cells at the tissue surface (Fig.~\ref{figCell}a). This situation corresponds to new bone formation by osteoblasts \emph{in vivo}, but also models \emph{in-vitro} neotissue growth in bioengineering scaffolds~\cite{Rumpler2008, Bidan2012, Bidan2013, Bidan2013b,Guyot2014,Guyot2015} where new tissue is predominantly produced near the tissue surface~\cite{Bidan2016}. It may also describe the growth of spheroid tumours that have proliferative outer rims~\cite{Lowengrub2010}, and wound healing~\cite{Rolli2012}. The normal velocity of the tissue surface is given by
\begin{align} 
	v = \kform \ \rho,   \label{eqnV}
\end{align}
where $\rho$ is the surface density of tissue-synthesising cells (number of cells per unit surface), and $\kform$ is the cells' secretory rate (volume of new tissue formed per cell per unit time)~\cite{Buenzli2015}. %  \cite{Buenzli2014a, Jones1974, Marotti1976, Volpi1981}.  
We assume here that tissue secretion is such that it displaces the cells perpendicularly to the surface at all times, i.e., cells are advected with velocity $\b v = v \b n$, where $\b n$ is the outward unit normal vector of the tissue surface.

We restrict in this paper to two spatial dimensions as we will compare our model to experimental data obtained from cross-sectional slices. We track the evolution of the tissue interface $S(t)$ by an explicit one-dimensional parameterisation $s\mapsto \b\gamma(s,t)$ of $S(t)$. Since the normal velocity of the tissue interface is given by $v$ in Eq.~\eqref{eqnV}, $\b\gamma$ must be such that the normal component of $\b\gamma_t$ matches $v$, i.e.,
\begin{align}\label{gamma_t}
    \b \gamma_t \cdot \b n = v,
\end{align}
where $\b n$ is the outward unit vector perpendicular to the tangential vector $\b \tau = \b\gamma_s/|\b\gamma_s|$, see Fig.~\ref{figCell}b.
(Partial derivatives are denoted by subscripts throughout the paper.) Since tissue geometry is unaffected by tangential components of interface velocity, Eq.~\eqref{gamma_t} is the only constraint that $\b\gamma$ must satisfy. In particular, we do not assume that the paths $t\mapsto \b\gamma(s,t)$ follow cell trajectories normal to $S(t)$ at each time. Later, $\b\gamma$ will be represented by tissue tickness functions in Cartesian and polar coordinates for which the tangential component $\b \gamma_t \cdot \b \tau \neq 0$.

On concave portions of the tissue substrate, new tissue deposition reduces the local surface area and thereby tends to increase cell density. By Eq.~\eqref{eqnV}, this leads to crowding of new tissue produced. On convex portions of the tissue substrate, new tissue deposition increases the local surface area and thereby tends to decrease cell density (Fig.~\ref{figCell}a). This leads to dispersion of new tissue produced. To describe this influence of local curvature on the evolution of cell density and tissue growth rate, we write $\rho$ at coordinate $s$ of the surface and time $t$ as
\begin{align}
	\rho(s,t) = \frac{\deltaup N(s,t)}{\deltaup \ell(s,t)}, \label{eqnRho}
\end{align}
where $\deltaup N$ is the number of cells residing on an infinitesimal length element $\deltaup \ell = g \der s$ centred at $s$, and $g=|{\b \gamma}_s|$ is the metric associated with $\b \gamma$~\cite{Kuehnel2006}. We then determine changes in cell density along the normal trajectories taken by the cells. The rate of change in $\rho$ in the normal direction is given by:
\begin{align}
	\dperp{}{t}\rho &\equiv \lim_{\Delta t \rightarrow 0} \frac{1}{\Delta t} \left[\rho(s+\Delta s ,t+\Delta t) - \rho(s,t)\right] \notag
    \\& = \rho_t - \frac{\b\gamma_t\cdot\b\tau}{g}\rho_s , \label{rho_t1}
\end{align}
where the $s$-coordinate offset $\Delta s$ is due to the fact that timelines $t\mapsto \b\gamma(s,t)$ at fixed $s$ are not normal to $S(t)$ in general. This offset is defined such that $v\Delta t\b n = \b\gamma_t \Delta t + \b\gamma_s \Delta s$ as $\Delta t \to 0$ (see Fig.~\ref{figCell}b). Projecting onto the tangential vector $\b\tau$ shows that $\lim_{\Delta t\to 0}\Delta s/\Delta t = -(\b\gamma_t\cdot\b\tau)/g$, which is used for the second equality in Eq.~\eqref{rho_t1}. %This also shows that Eq.~\eqref{eqnV} is equivalent to $\b\gamma_t = v\b n + \b\tau(\b\gamma_t\cdot\b\tau)$
The differential operator $\dperp{}{t}$ corresponds here to the substantial derivative that follows the advective velocity~$v\b n$. This operator obeys standard differentiation rules such that with Eq.~\eqref{eqnRho}, one has
\begin{align}
	\dperp{}{t} \rho     
	\ = \  \frac{\dperp{}{t} \deltaup N}{\deltaup \ell}\ -\ \rho \frac{\dperp{}{t} \deltaup \ell}{\deltaup \ell}. \label{rho_t2}
\end{align}
The second term in the right hand side of Eq.~\eqref{rho_t2} represents a geometric contribution to density changes due to changes in the local length of the surface $\deltaup \ell$ induced by the surface's evolution. This contribution is related to the tissue substrate curvature $\kappa(s,t)$ by:
\begin{align}\label{stretch}
    \frac{\dperp{}{t} \deltaup \ell}{\deltaup \ell} = v \kappa
\end{align}
(see Appendix~\ref{appx:stretch}). We use the convention that $\kappa<0$ on concave portions of the substrate and $\kappa >0$ on convex portions of the substrate. %\todo{happy to swap} {\color{blue}  I thought we would follow Bidan's etc.}
The first term in the right hand side of Eq.~\eqref{rho_t2} represents a contribution to density changes due to changes in the number of cells $\deltaup N$ populating the length element $\deltaup \ell$. These changes may occur by nonconservative and conservative processes, such as cell creation, cell elimination, and cell transport along the surface. We assume here that $\deltaup N$ changes due to cells being eliminated from the active pool at rate $A$ (probability per unit time) and diffusing along the bone surface with constant diffusivity $D$, giving
\begin{align}
	\frac{\dperp{}{t} \deltaup N}{\deltaup \ell} = D \rho_{\ell \ell} - A \rho,   \label{cellnum_l} 
\end{align} 
where $\partial / \partial \ell = (1/g)\partial / \partial s$ is the partial derivative with respect to the arc length $\ell$ ($\pd{^2}{\ell^2}=g^{-2}\pd{^2}{s^2} - g^{-3}\b\tau\cdot\b\gamma_{ss}\pd{}{s}$ is the one-dimensional Laplace--Beltrami operator~\cite{Berger2003}). Combining Eqs~\eqref{rho_t1}--\eqref{cellnum_l}, the evolution of cell density is governed by
\begin{align}
	\rho_t & =  \frac{\b\gamma_t\cdot\b\tau}{g} \rho_s
	- \rho v \kappa    
	+ D \rho_{\ell \ell} 
	- A\rho.  \label{rho_t3}
\end{align}
The first term in the right hand side of Eq.~\eqref{rho_t3} depends on the choice of parameterisation $\b\gamma$. It describes the transport of cells normal to the interface measured with respect to the coordinate $s$. It is absent if $\b\gamma$ is chosen to be an orthogonal parameterisation, defined such that $\b\gamma_t\cdot\b\gamma_s=0$. The second term represents the systematic dilution or concentration of cell density induced by the (signed) curvature of the interface. The third and fourth terms describe the diffusion of cells parallel to the interface, and the cell depletion rate, respectively.

Equation~\eqref{rho_t3} is coupled to the evolution of the tissue surface $S(t)$ via Eqs~\eqref{eqnV} and~\eqref{gamma_t}. Notice that since $\rho$ is proportional to the normal velocity of $S(t)$, the contribution $-\rho v\kappa = -\kform \rho^2\kappa$ to $\rho_t$ in Eq.~\eqref{rho_t3} implies that the normal acceleration of the surface depends linearly on curvature, which constitutes a type of hyperbolic curvature flow~\cite{LeFloch2008}. This is to be contrasted with mean curvature flow in which the normal velocity depends linearly on curvature~\cite{Grayson1987,Sethian1999}. The nonlinearity of the equations and their hyperbolic character suggest that shocks may develop, e.g. as cusps in the interface $S(t)$. This situation requires to seek weak solutions, such as entropic solutions found by adding infinitesimal diffusion of the interface, or equivalently, by devising diffusive (e.g., upwind) numerical schemes~\cite{Crandall1983,Lax1973,Harten1983,LeVeque2004}. In our case, a physiologically relevant weak solution additionally requires that cell densities remain finite at developing cusps of the interface. This is ensured by the explicit inclusion of cell diffusion along the interface. We note here that radii of curvature $1/\kappa$ of the order of a single cell size ($\approx 20$--$30\,\um$) may be considered cusps already within the continuum model. However, weak, entropic solutions provide a physically consistent extension of the continuum model below such radii of curvature.

In summary, the systematic effect of curvature onto cell density is expected to help smooth substrate irregularities by generating a curvature-dependent normal acceleration, while active cell depletion is expected to capture tissue deposition slowdown.

\paragraph{Scaling analysis and choice of units} \hspace{-2mm}The mathematical model involves five generic parameters: a characteristic length scale of the initial substrate geometry $\b\gamma(s,0)$ (e.g., a radius of curvature $R_0$); a characteristic value $\rho_0$ of the initial cell density $\rho(s,0)$; a characteristic value $\kform^0$ of the secretory rate $\kform(s,t)$; the diffusivity $D$; and the cell depletion rate $A$. Through a scaling analysis in which cell density, space, and time are rescaled, it is possible to show that only two of these five parameters are independent. We choose these parameters to be the diffusivity $D$ and the cell elimination rate $A$ without restriction of generality. We can thus fix arbitrarily the length scale of the initial substrate, the parameter $\rho_0$, and the parameter $\kform^0$, and explore the qualitative behaviours of the solutions by modifying only $A$ and $D$.

In Section~\ref{Sect4b}, the length scale will be set to match the physical size of the experimental initial substrates (in mm), and the product $v_0 = \kform^0\rho_0$ will be set to match the experimental initial normal velocity (in mm/\da). This ensures that $D$ (in $\mm^2/\da$) and $A$ (in $\da^{-1}$) have proper physical dimensions. While the normal velocity is easily deduced experimentally, cell density (in $\mm^{-2}$) and secretory rate (in $\mm^3/\da$) are difficult to estimate, and they are usually not measured. In bone \emph{in vivo}, the density of osteoblasts ranges from about \num{2000}--\num{10000}\,$\mm^{-2}$ (see~\cite{Buenzli2014a} and Refs cited therein). In the \emph{in-vitro} bioscaffold tissue growth experiments of Refs~\cite{Rumpler2008, Bidan2012, Bidan2013}, the seeding density is 800--1000\,$\mm^{-2}$, but the initial confluent density at the onset of formation is not known. The evolution of the tissue interface does not actually depend on the relative proportion of cell density and secretory rate in~$v$ in Eq.~\eqref{eqnV}. For ease of interpretation, we will thus choose in the remainder of the paper units in which $\kform^0=1$ is dimensionless, so that $\rho$ corresponds to $v$ (in $\mm/\da$) by Eq.~\eqref{eqnV}. This scaling only affects the units of $\rho$ and $\kform$.\footnote{This is equivalent to first considering the scaled density $\overline\rho\equiv \kform^0 \rho$ and scaled secretory rate $\overline{\kform^0} \equiv \kform^0/\kform^0$ where $\kform^0$ has units, and then dropping the bars from the notation.}

\paragraph{Conservative form and total cell number}
Numerical simulations of direct discretisations of Eq.~\eqref{rho_t3} using finite difference upwind schemes were found to induce significant numerical nonconservation of the total number of cells at low diffusivities~$D$, due to developing cusps in the interface. For these situations, the equations were first rewritten in conservative form, and then discretised using finite volume conservative numerical schemes (see below).

The conservative form of Eq.~\eqref{rho_t3} is found for a general parameterisation $\b\gamma$ by considering the projected density of cells on the $s$ coordinate, $\eta(s,t)$, defined such that $\deltaup N = \eta \der s$ is the number of cells on the interface between the coordinates $s$ and $s+\der s$. Since $\deltaup N = \rho \deltaup\ell = \rho g \der s$, one has $\eta(s,t) = \rho(s,t)g(s,t)$. It is shown in Appendix~\ref{appx:conservative} that
\begin{align}
	\eta_t + \left[-\frac{\eta}{g}({\b \gamma}_t \cdot \b{\tau})  
	- \frac{D}{g} \left(\frac{\eta}{g} \right)_s\right]_s = - A\eta   \label{eta_t}.
\end{align}
Equation~\eqref{eta_t} is a conservation law that expresses the balance of cells between $s$ and $s+\der s$ during the evolution. For periodic or no-flux boundary conditions, the total number of cells on the whole interface $N(t) \equiv \int_a^b\der s\  \eta(s,t)$ evolves as
\begin{align} \label{cellnum_t}
    \td{N}{t} = \int_a^b\hspace{-1ex}\der s\ \eta_t = - \int_a^b\hspace{-1ex}\der s\ A \eta,
\end{align}
since the contribution to the integral of the flux term of Eq.~\eqref{eta_t} is zero. If the cell elimination rate~$A$ is homogeneous, $N(t) = N_0 \exp\Big(-\int_0^t \der t\, A\Big)$, as expected.

\begin{figure*}[t]
% \captionsetup{justification=centering}

\centerline{\includegraphics[trim={10 50 30 0}, width=1\linewidth,clip]{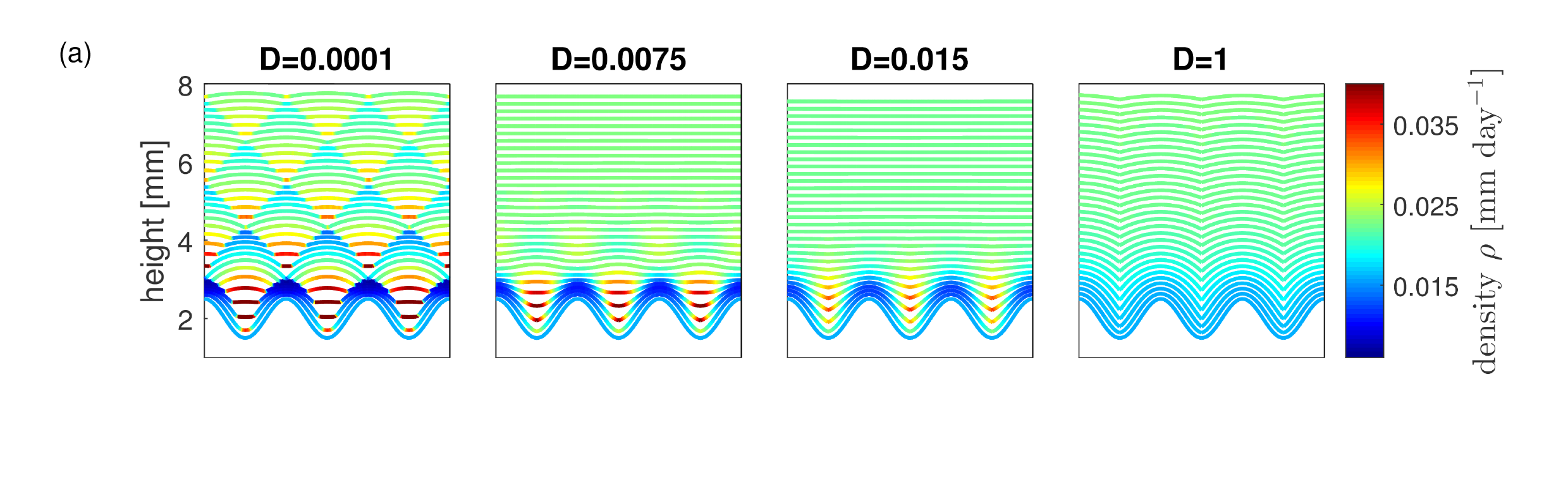}}
\centerline{\includegraphics[trim={10 27 30 30}, width=1\linewidth,clip]{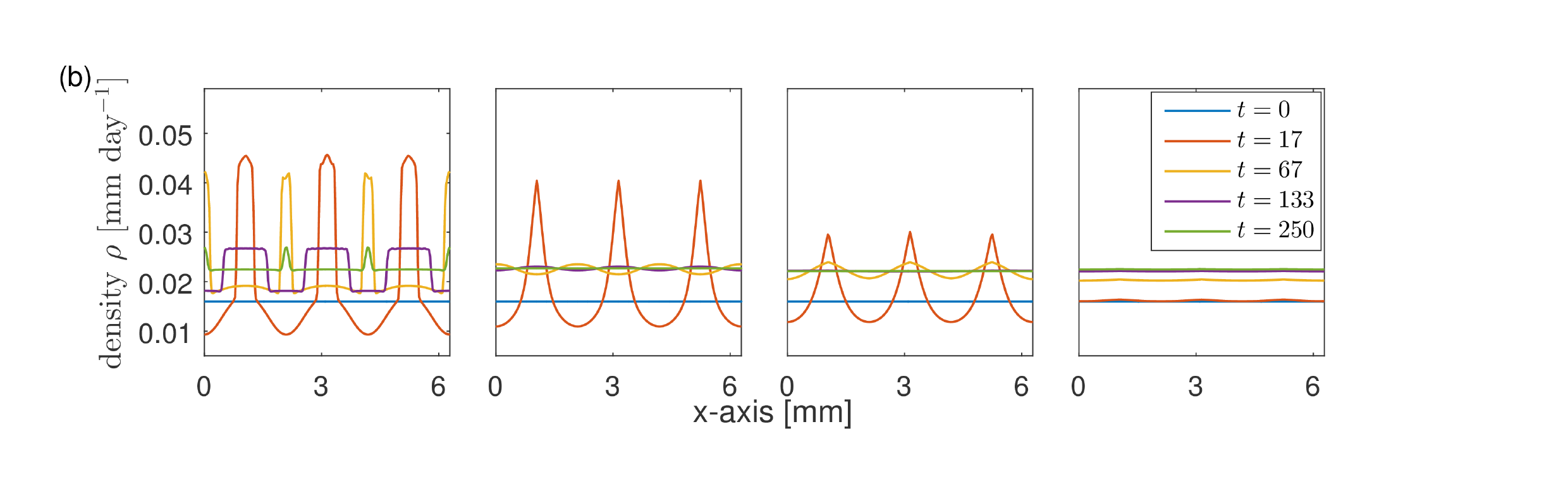}}

\caption{Tissue deposition on a cosine initial substrate for a range of low to high diffusivities ($D=0.0001$, $0.0075$, $0.015$, $1$\,$\mm^2/\da$). (a) Evolution of the tissue interface. Each line corresponds to the interface $h(x,t)$ at regular time intervals $\Delta t=8.33\,\days$, and is coloured according to cell density. (b) Cell density profiles at specific times. Simulations performed with $A=0$, $\kform^0=1$, and $\rho_0=0.016\,\mm/\da$.}
\label{figCosine}
\end{figure*}

\paragraph{Numerical discretisation}
At high cell diffusivity, we used a straightforward semi-implicit finite difference discretisation of the equations for $\rho$ and $\b\gamma$. First-order and second-order spatial derivatives were discretised using upwind and central differencing, respectively. Advective and reaction terms were solved explicitly with forward Euler discretisation in time, while diffusive terms were solved implicitly with backward discretisation.

At low cell diffusivity, this finite difference scheme led to numerical nonconservation of cells requiring finer space grid resolution. To prevent the numerical nonconservation of cells, we discretised the conservative form of the equations with the finite volume method instead~\cite{LeVeque2004}. We implemented the semi-discrete Kurganov-Tadmor scheme~\cite{Kurganov2000} with a fully explicit forward Euler discretisation in time.

Both numerical schemes give indistinguishable results in a range of intermediate diffusivities. The maximum numerical error on cell number recorded in all our simulations was 3\% (triangular pore, $D=0.005$). All other simulations had less than 1\% cell number error. More details on these numerical schemes can be found in Appendix~\ref{appx:num}.

\section{Applications and numerical results}
\label{Sect4}
During bone remodelling, new bone formation occurs on various types of bone interface topologies. In porous, meshed trabecular bone, new bone tissue is deposited on the floor of trench-like cavities of zero average curvature carved out of single struts. In dense cortical bone, new bone tissue is deposited on the walls of porous channels~\cite{Martin2000}. Neotissue deposition in porous bioscaffolds has also been investigated on trench-like cavities or within channels of various cross-sectional shapes~\cite{Rumpler2008, Bidan2012, Bidan2013, Bidan2013b}.

We apply our mathematical model to these two classes of surface topologies by parameterising $S(t)$ with thickness functions in Cartesian or polar coordinates, respectively. Tissue deposition in trench-like cavities of zero average curvature is represented by an evolving height $y=h(x,t)$ with periodic boundary conditions. Tissue deposition in porous channels is represented by an evolving radius $r = R(\theta, t)$. Both $h(x,t)$ and $R(\theta,t)$ represent the local thickness of newly-deposited tissue material at constant value of the parameter $s=x$ in Cartesian coordinates, $s=\theta$ in polar coordinates. The governing equations for $\b\gamma$ and $\rho$ (or $\eta$) are specialised to these non-orthogonal parameterisations of $S(t)$ (Appendix~\ref{appx:gov-eq}), discretised, and solved numerically (Appendix~\ref{appx:num}).

\subsection{Influence of cell diffusion on interface smoothing}
\label{Sect4a}
We start by investigating the smoothing of an initially rugged substrate due to the volumetric crowding of tissue modelled by the hyperbolic curvature flow proposed in this paper. We first assume that cells are not eliminated, i.e., $A=0$, and that they produce new tissue at a constant rate $\kform=\kform^0=1$ (dimensionless, so that $\rho$ corresponds to $v$ by Eq.~\eqref{eqnV}, see Sec.~\ref{Sect2}). Since interface smoothing can be expected to depend significantly on the amount of cell diffusion parallel to the interface~\cite{Simpson2006}, we performed simulations using a range of diffusivities $D$ both in trench-like cavities and in porous channels.

Figure~\ref{figCosine} shows the evolution of a trench-like initial interface with a rugosity modelled by cosine oscillations:
\begin{align}  \label{height0}
    h(x,0) = 2 + \tfrac{1}{2}\cos(3 x), \qquad x\in [0,2\pi).
\end{align}
The surface is initially seeded with a homogenous cell density $\rho(x,0) = \rho_0 = 0.016\,\mm/\da$ (this value is calibrated from the pore scaffold tissue growth experiments of Ref.~\cite{Bidan2012}, see Sec.~\ref{Sect4b}). The evolution is shown for different cell diffusivities $D$. Coloured lines in Fig.~\ref{figCosine}a represent the interface $h(x,t)$ at regular time intervals $\Delta t=8.33\,\days$ starting from $t=0$. These interfaces are coloured by the corresponding cell density $\rho(x,t)$. Plots of cell densities are also shown at specific times in Fig.~\ref{figCosine}b.

At low diffusivity ($D=0.0001\,\mm^2/\da$), concave portions of the interface rapidly concentrate cells (red), which increases the local propagation speed, while convex portions of the interface disperse cells (dark blue), which decreases the local propagation speed. The stark contrast in local propagation speed generates cusps in the interface, that propagate sideways as shock waves between the concave and convex regions. These shock waves collide and bounce off each other, resulting in an oscillatory spatio-temporal pattern whereby concave portions of the interface become convex, and convex regions of the interface become concave repetitively. With increasing diffusivity, cusps in the interface smooth out and this oscillatory pattern dampens more rapidly ($D=0.0075\,\mm^2/\da$). At $D=0.015\,\mm^2/\da$, the interface smoothens to a flat interface without oscillatory pattern the quickest (see below). At large diffusivities ($D\gtrsim 1\,\mm^2/\da$), cell concentration and dispersion effects are entirely overridden by the diffusive redistribution of cells, resulting in nearly homogeneous cell densities throughout the simulation. The interface evolves by constant offsets in the normal directions. The size of these offsets increases with time because the total length of the interface decreases and therefore, the overall cell density increases. This kind of evolution by normal offsets is well-known to creates cusps in the interface within a finite range. For constant normal velocity, these cusps disappear at rate $\Order\big(t^{-1}\big)$ as $t\to\infty$~\cite{Kardar1986}. A faster rate of cusp disappearance occurs in our case as normal velocity is linked to total interface length.

\begin{figure}
% \captionsetup{justification=centering}

\centerline{
	\includegraphics[trim={0 0 0 0}, width=1\linewidth,clip]{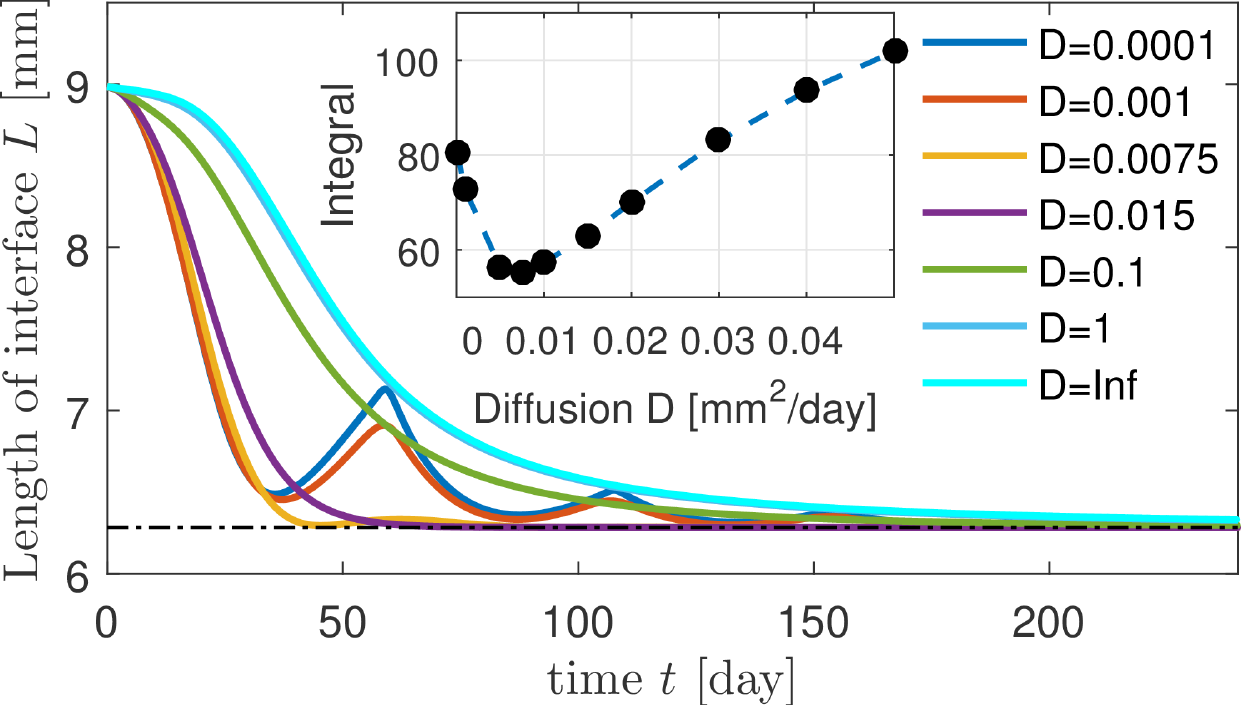}}

\caption{Influence of diffusivity on the rate and manner of smoothing of an initial cosine interface. The total length of the interface transitions from damped oscillation regimens at low diffusivities, to critically damped regimens at intermediate diffusivities, to overdamped regimens at high diffusivity. \emph{Inset}: the minimum integral of the timeline $t\mapsto L(t)$ is reached at a critical diffusivity $D\approx 0.0075\,\mm^2/\da$ smaller than the critical diffusivity $D\approx 0.015\,\mm^2/\da$ at which oscillating patterns are lost.}
\label{figLength}
\end{figure}

\begin{figure*}[t]
% \captionsetup{justification=centering}

\centerline{
\includegraphics[trim={30 52 30 20 },width=0.8\linewidth,clip]{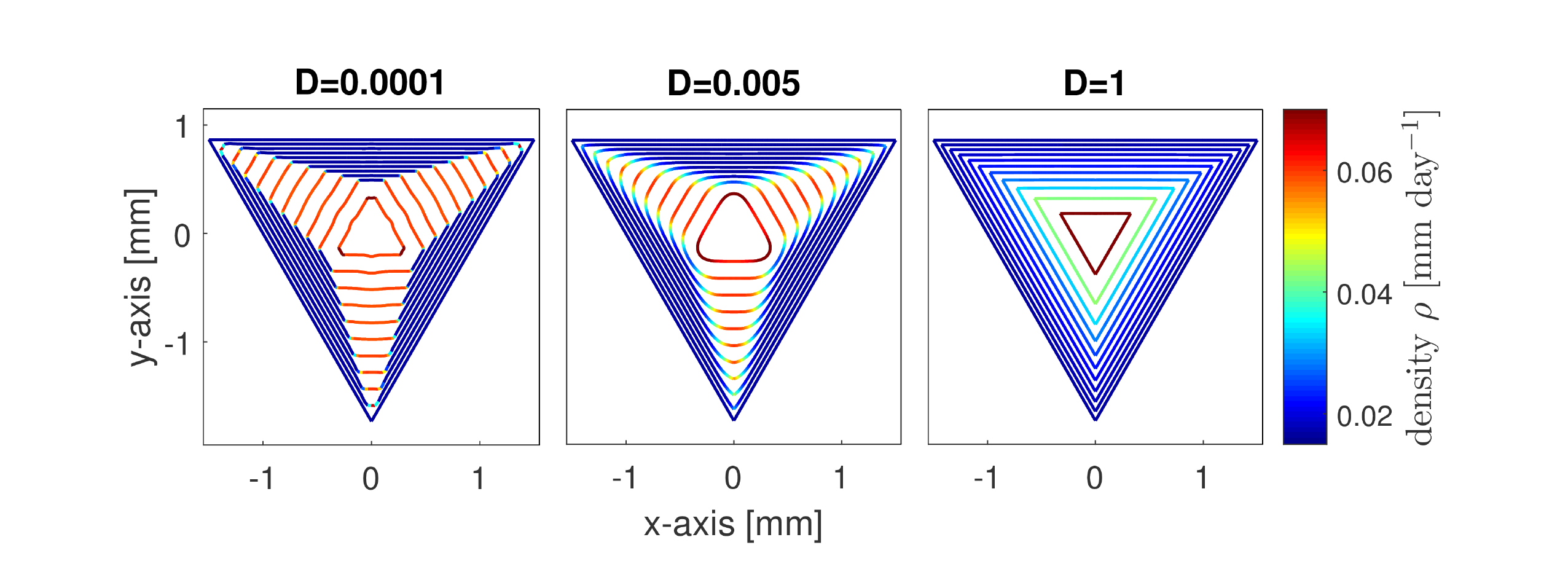} 
}

\centerline{
\includegraphics[trim={30 55 30 50} ,width=0.8\linewidth,clip]{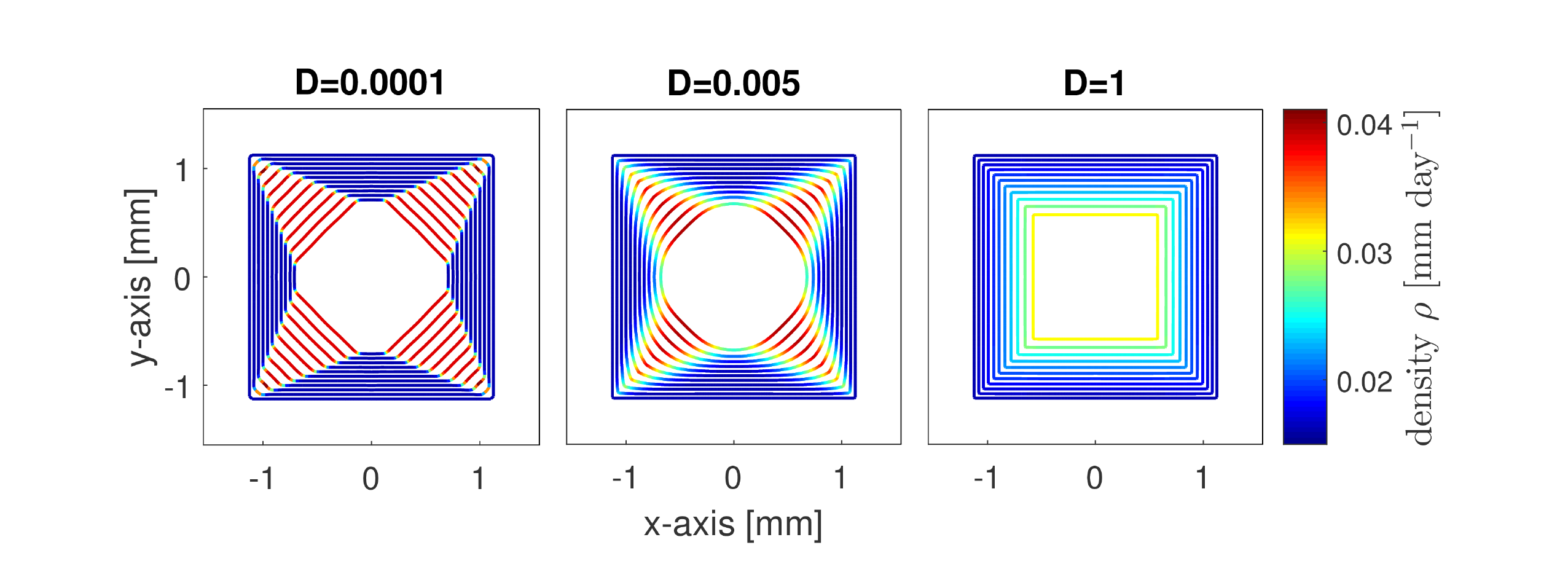} 
}

\centerline{
\includegraphics[trim={30 5 30 47},width=0.8\linewidth,clip]{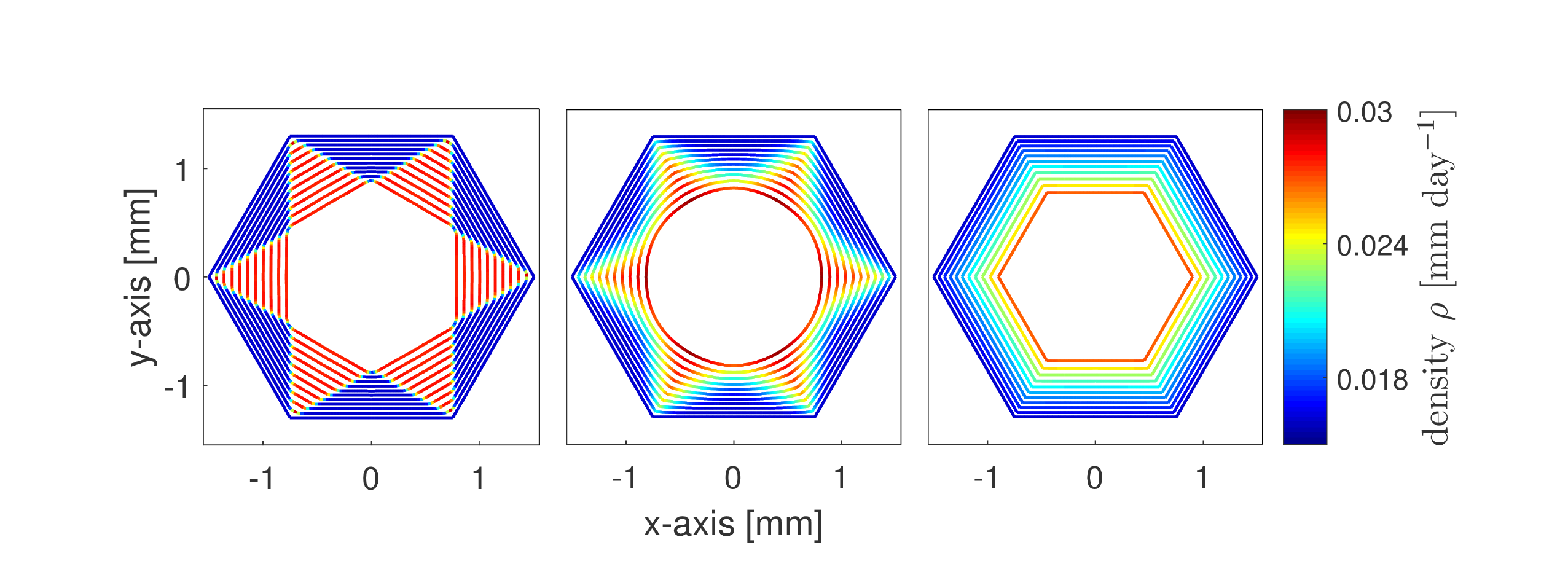}
}

\caption{Tissue deposition within triangular, square and hexagonal pores (each with initial perimeter $9\,\mm$) for low ($D=0.0001\,\mm^2/\da$), intermediate ($D=0.005\,\mm^2/\da$), and high ($D=1\,\mm^2/\da$) diffusivities. The tissue interface is shown at regular time intervals $\Delta t=2.6\,\days$ until $t=26\,\days$ and coloured according to cell density. Simulations performed with $A=0$, $\kform=1$, $\rho_0=0.016\,\mm/\da$.}
\label{figVariousShapes}
\end{figure*}
It is clear from Figure~\ref{figCosine} that the diffusivity $D$ drives strong qualitative changes in the evolution of the interface, that influence in particular the rate of interface smoothing. The interface's total length $L(t)=\int_0^{2\pi}\der x\, g(x,t)$ in Figure~\ref{figLength} converges to the minimum length $2\pi$ (flat interface) by transitioning from damped oscillation regimens at low diffusivity, to critically damped regimens at intermediate diffusivities, and to overdamped regimens at high diffusivity. The situation is similar to a damped harmonic oscillator except that two critical diffusivities can be distinguished: one for which the integral of the timelines $t\mapsto L(t)$ is minimal ($D\approx 0.0075\,\mm^2/\da$, yellow curve; see also inset); and one above which oscillating interface patterns do not occur ($D\approx 0.015\,\mm^2/\da$, purple; see also Fig.~\ref{figCosine}). In a critically damped harmonic oscillator, these two critical behaviours coincide~\cite{Serway2007}.

The strength of diffusivity $D$ drives similar qualitative changes in the evolution of porous channels during tissue deposition (Figure~\ref{figVariousShapes}). At low diffusivity ($D=0.0001\,\mm^2/\da$), the curvature-induced increase in cell density and resulting tissue crowding at corners of the initial pore shape increases the local propagation speed of the interface (red). New cusps in the interface are created laterally due to the contrast in propagation speed. These cusps propagate sideways as shock waves and collide. At intermediate diffusivity ($D= 0.005\,\mm^2/\da$), cusps smooth out and the interface develops into a circular shape (at a rate that depends on acuteness). At high diffusivity ($D = 1\,\mm^2/\da$), cell density is homogeneous, but increases with time as the interface's total length decreases. Initial cusps in the interface are maintained throughout the evolution.

\subsection{Application to bioscaffold tissue growth}\label{Sect4b}
We now apply our mathematical model to the \emph{in-vitro} experiments of Refs~\cite{Rumpler2008, Bidan2012, Bidan2013, Bidan2013b} in which tissue was grown on bioscaffolds of various shapes. In these experiments, hydroxyapatite bioscaffolds were initially seeded with a uniform density of cells. However, no tissue was produced on convex portions of these substrates. This suggests that the secretory rate $\kform$ in Eq.~\eqref{eqnV} is itself a function of curvature, such that no tissue matrix is secreted by the cells when $\kappa \geq 0$. We take this function to be:
\begin{align}\label{kform2}
    \kform(\kappa) = \begin{cases} \kform^{0}, &\quad \text{if $\kappa < 0$},
\\ 0, &\quad \text{if $\kappa \geq 0$},
\end{cases}
\end{align}
where $\kform^0$ is a constant. With Eq.~\eqref{kform2}, the normal velocity of the interface is zero on convex portions of the interface, and it accelerates in proportion to curvature on concave portions of the interface as per Eq.~\eqref{rho_t3}. In Refs~\cite{Rumpler2008,Bidan2012,Bidan2013,Bidan2013b}, the authors suggested the phenomenological model of tissue growth given by $v= - \lambda \kappa$ if $\kappa<0$, and $v=0$ if $\kappa \geq 0$. With this phenomenological model the total cross-sectional area $A_\text{T}(t)$ of new tissue produced up to time~$t$ increases at constant rate on pore substrates that are concave everywhere. Indeed,
$A_\text{T}'(t) \equiv \int_0^{P(t)}\der \ell v(\ell,t) = 2\pi \lambda$ by the total absolute curvature theorem, where $\ell$ is the arc length and $P(t)$ is the pore's perimeter~\cite{Kuehnel2006}. This was used with experimental determinations of $A_\text{T}(t)$ in circular pore shapes~\cite{Bidan2012} and square pore shapes~\cite{Bidan2013} to calibrate $\lambda$. Because rates of tissue growth $A_\text{T}'(t)$ decreased at large times in the experiments (indicating tissue formation slowdown), this calibration was performed at the onset of tissue growth, assumed here to be $t=0$, by setting $\lambda = A_\text{T}'(0)/(2\pi)$.

\begin{figure}[!t]
% \captionsetup{justification=centering}

\hspace*{-1.2mm}\centerline{
\includegraphics[%trim={15 0 100 0}, 
width=1.04\linewidth]{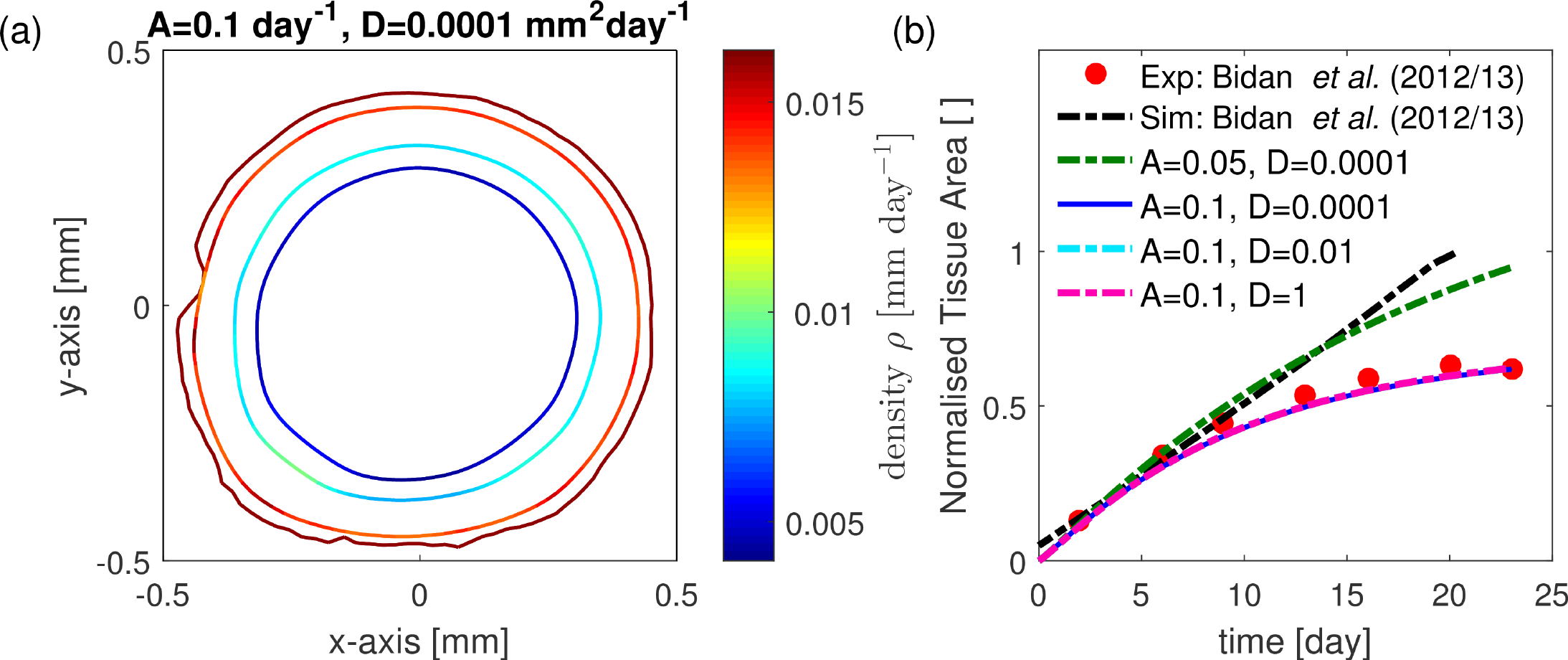}
}

\hspace{0mm}\centerline{
\includegraphics[%trim={20 0 100 0}, 
width=1\linewidth,clip]{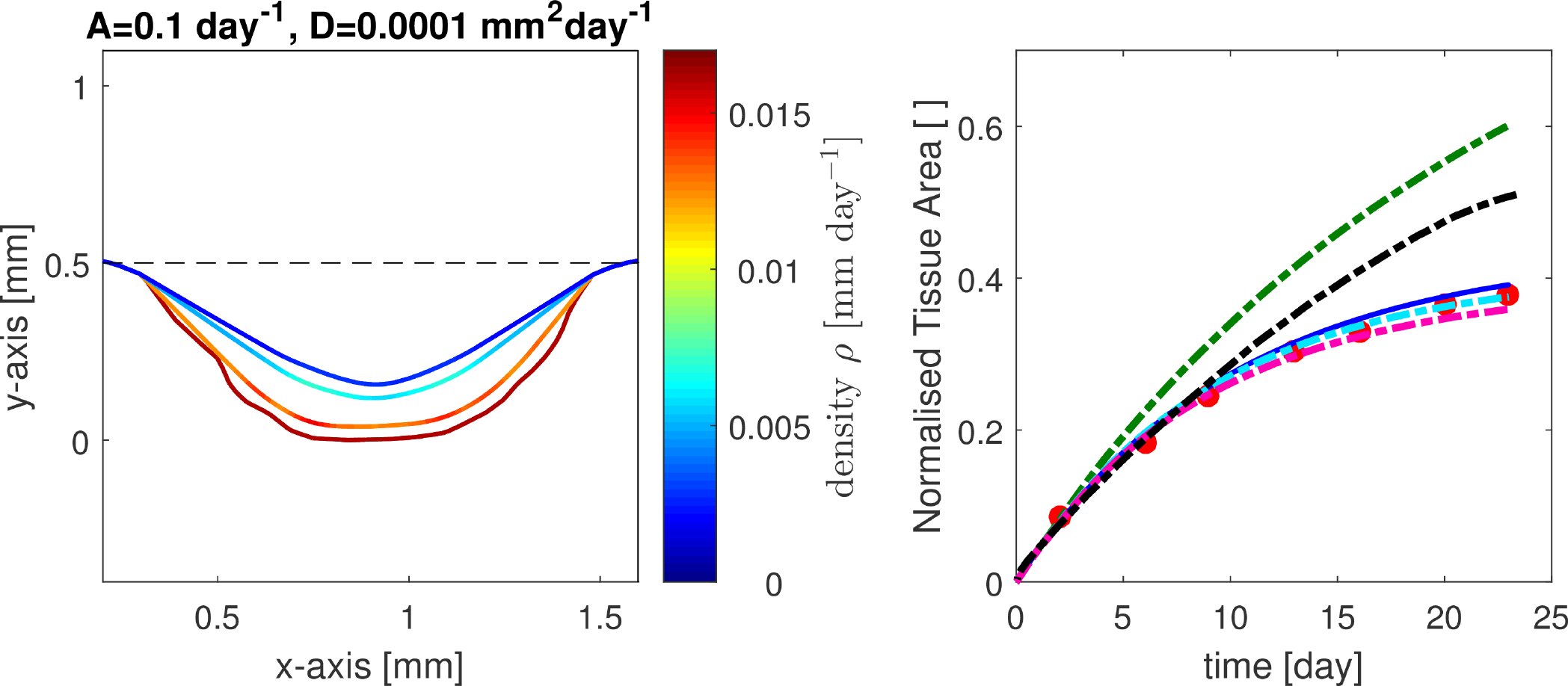}}

\centerline{
\hspace{0mm}\includegraphics[%trim={20 0 100 0}, 
width=1\linewidth,clip]{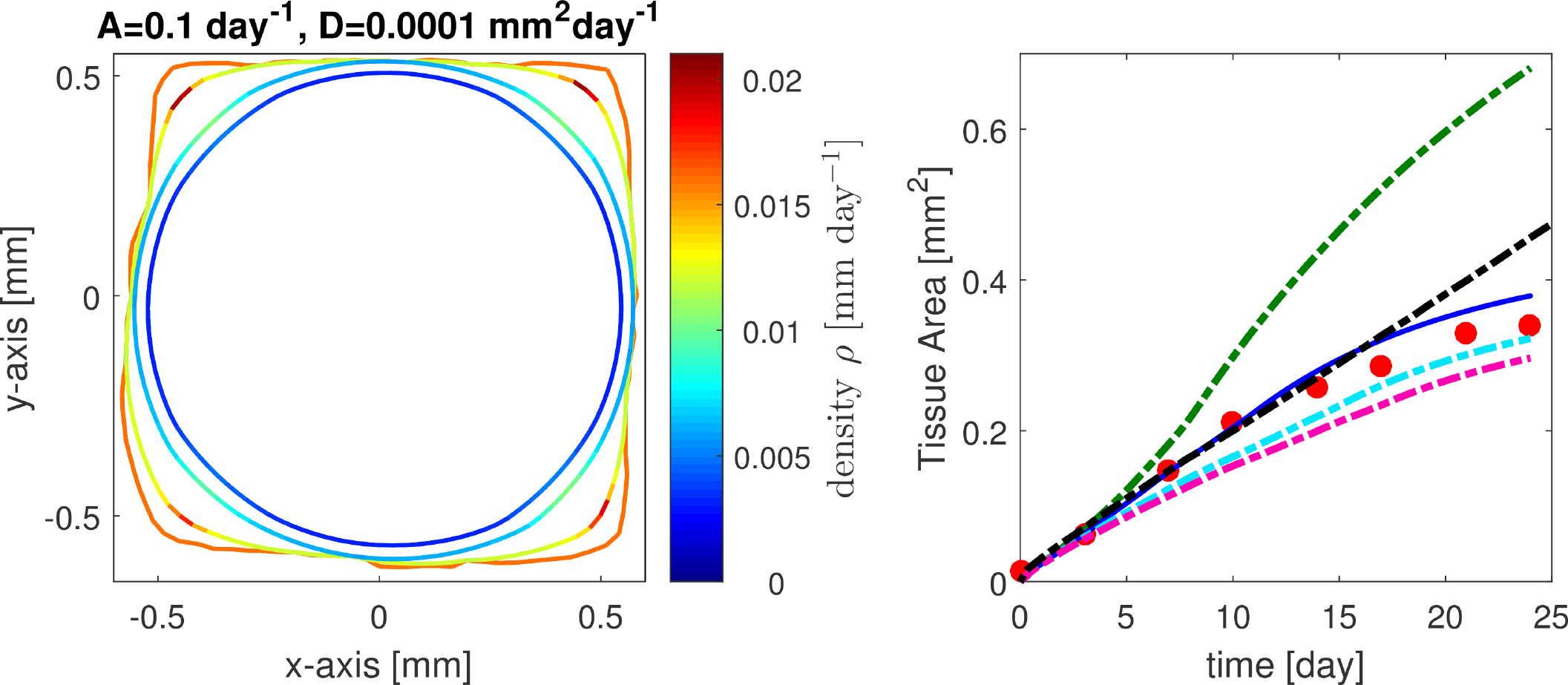}}

\centerline{
\hspace{0mm}\includegraphics[%trim={20 0 100 0}, 
width=1\linewidth,clip]{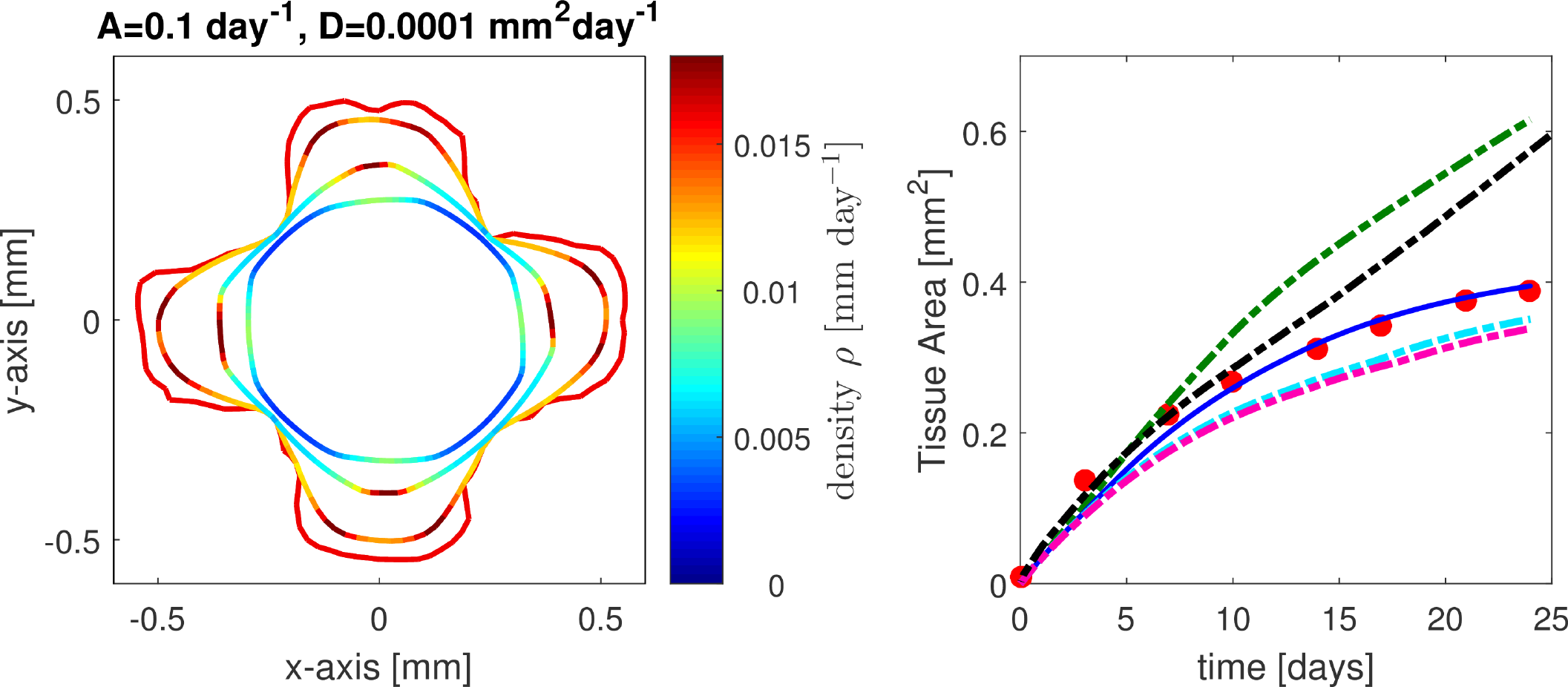}}

\caption{Tissue deposition predicted by our cell-based model with Eq.~\eqref{kform2} in the bioscaffold pore shapes of Bidan~\etal~\cite{Bidan2012} (circular and semi-circular pore shapes) and Bidan~\etal~\cite{Bidan2013} (square and cross pore shapes). \emph{(a)} The tissue interface is shown at days 4, 7, 14, 21 and colored according to cell density. \emph{(b)} The time evolution of the total tissue area produced $A_T(t)$ (normalised by the initial pore area in the circular and semi-circular cases) is shown for various values of diffusivity $D$ and cell elimination rate $A$. These time evolutions are compared with the experimental results and simulations of the phenomenological model of Refs~\cite{Bidan2012,Bidan2013}.}\label{figCompareBidan}
\end{figure}

Our cell-based model is equivalent to this phenomenological model when tissue is deposited within perfectly circular pores and cells are not eliminated ($A=0$). Indeed, in this instance, by Eqs~\eqref{eqnV} and~\eqref{kform2}:
\begin{align}\label{mcf-circle}
v = \kform \rho = - \frac{\kform^0N_0}{2\pi}\kappa \equiv - \lambda\kappa,
\end{align}
since $\rho = N_0/(2\pi R)$ and $\kappa = - 1/R$, where $R$ is the pore radius, and $N_0$ is the initial number of cells lining the circle's circumference. If active cells are depleted at constant rate $A$, the total number of active cells in our model decreases as $N(t)=N_0\e^{-At}$ and the proportionality coefficient $\lambda$ between velocity and curvature simply becomes time-dependent:
\begin{align}\label{calibration1}
    \lambda(t) = \frac{\kform^0 N_0\e^{-At}}{2\pi} = \kform^0 \rho_0\, R_0\,\e^{-At},
\end{align}
where $\rho_0=N_0/(2\pi R_0)$ is the initial cell density and $R_0$ is the initial pore radius. In non-circular pore geometries, our cell-based model does not reduce to mean curvature flow. However, under the assumption that $\kform^0$ and the initial seeding density $\rho_0$ are independent of initial pore shape, Eq.~\eqref{calibration1}, valid in the circular pore geometry, enables us to calibrate $v_0=\kform^0\rho_0$, the positive part of the initial normal velocity, in all pore geometries. From the experimental data $A_\text{T}(t)/(\pi R_0^2)$ in Figure~4B of Ref.~\cite{Bidan2012} ($R_0\approx 0.5\,\mm$), we estimate that the initial tissue production rate measured in circular pore scaffolds is $A_\text{T}'(0)\approx 0.051\,\mm^2/\da$.\footnote{The value $A_\text{T}'(0)\approx 0.0284\,\mm^2/\da$ reported by Bidan \etal~\cite{Bidan2012} should be corrected to $\approx 0.035\,\mm^2/\da$ to match the growth rate reported in their figure~4C from fitting linearly the first three experimental points of their figure~4B (the data is reproduced in Fig.~\ref{figCompareBidan}b, circular shape). Since our model accounts explicitly for tissue deposition slowdown, we estimated the slope at the onset of tissue growth by quadratic interpolation of the first five experimental points instead, giving a larger value.% This also gave an onset of tissue growth around 5\,days rather than 4\,days post seeding.
} We thus get from Eq.~\eqref{calibration1}:
\begin{align}\label{calibration2}
    v_0 = \kform^0\rho_0 = 
\frac{\lambda(0)}{R_0} = \frac{A_\text{T}'(0)}{2\pi R_0} \approx 16\,\um/\da.
\end{align}
As in Sec.~\ref{Sect4a}, for ease of interpretation, we choose units in which $\kform^0=1$ is dimensionless, so that $\rho$ corresponds to $v$ where $\kappa<0$. We set the initial (scaled) density $\rho_0 = v_0 = 0.016\,\mm/\da$ in all the numerical simulations.

Figure~\ref{figCompareBidan} shows the growth of new tissue predicted by our cell-based model with Eqs~\eqref{kform2} and~\eqref{calibration2} in the circular, semi-circular, square, and cross-shaped bioscaffold pores of Bidan~\etal~\cite{Bidan2012,Bidan2013}. The initial rate of tissue growth $A_\text{T}'(0)$ depends on $v_0$ and on the geometry of the initial substrate. While $v_0$ is calibrated from measurements of $A_\text{T}'(0)$ in the circular pore geometry, the rates $A_\text{T}'(0)$ obtained with the same value of $v_0$ in the other geometries (initial slope of the curves in Fig.~\ref{figCompareBidan}b) closely match the experimental initial growth rates. Remarkably, our cell-based model reproduces the experimental tissue growth curves $A_\text{T}(t)$ accurately---including tissue deposition slowdown---in all pore geometries for a single combination of diffusivity and cell depletion rate, $D=0.0001\,\mm^2/\da$ and $A=0.1/\da$. At these values, the interface rounds off efficiently regardless of initial pore shape, as observed experimentally~\cite{Rumpler2008,Bidan2012,Bidan2013} (see 
Fig.~\ref{figCompareBidan}a).

\begin{figure*}
% \captionsetup{justification=centering}

\centerline{
\includegraphics[trim={55 60 90 35} ,
width=0.9\linewidth,clip]{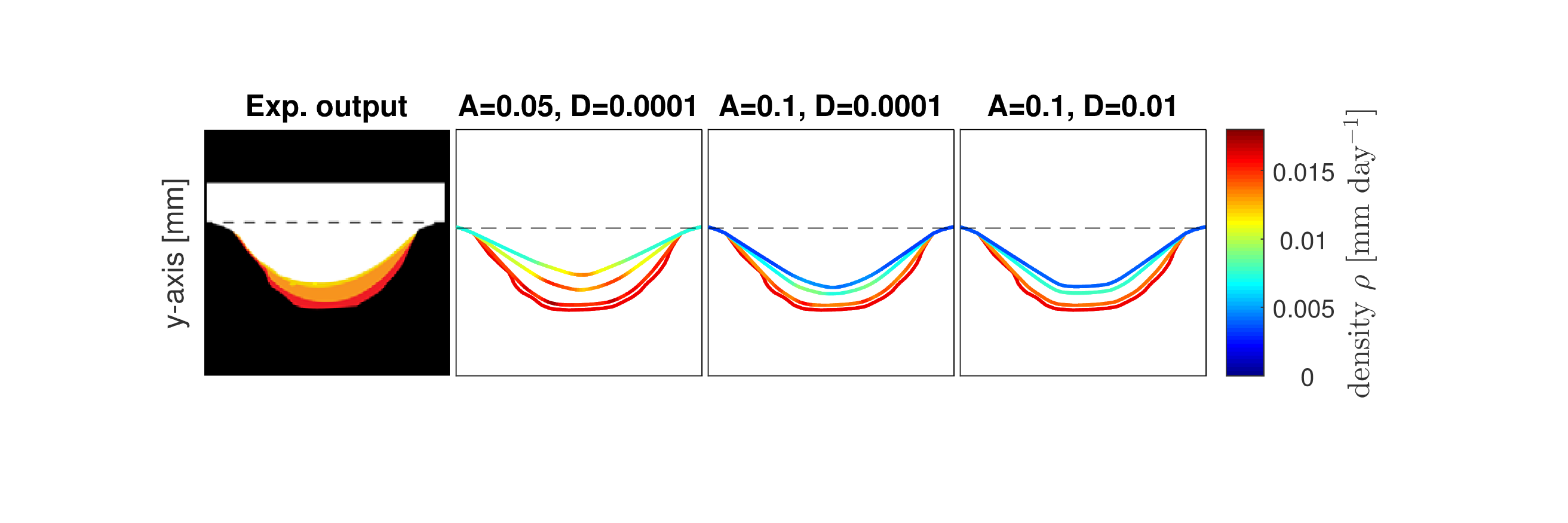}}
% \vskip-15mm
\centerline{
\includegraphics[trim={55 30 90 50} ,width=0.9\linewidth,clip]{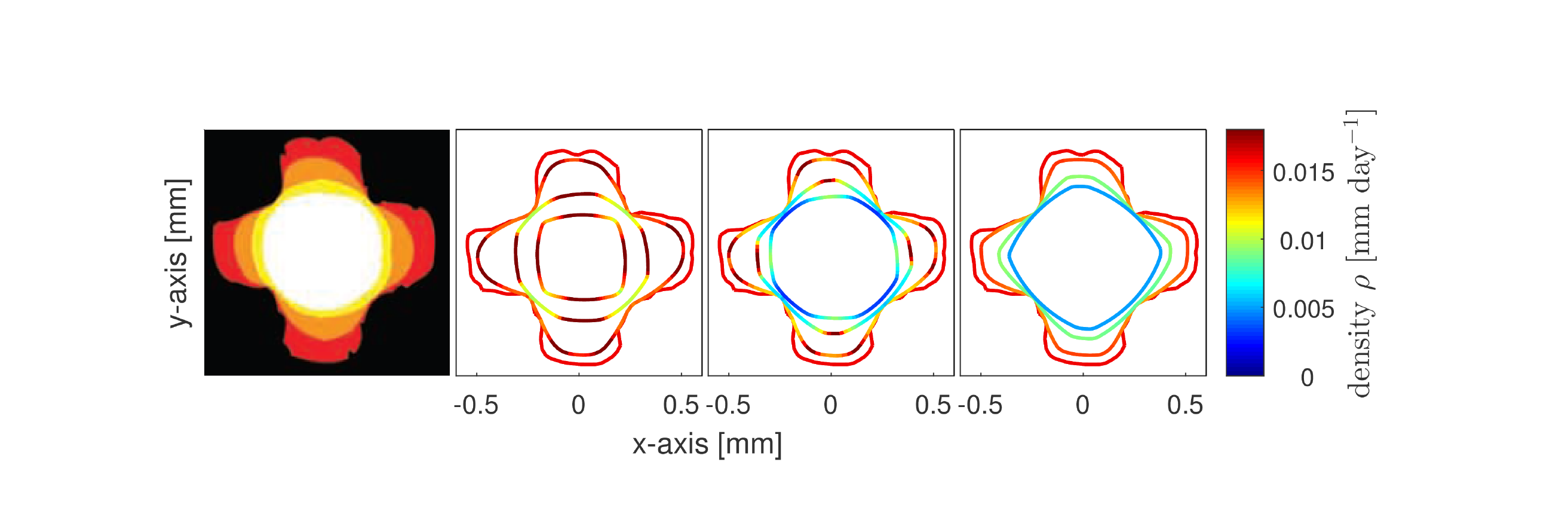}}

\caption{Evolution of the tissue interface of the cell-based model in the semi-circular (top) and cross (bottom) interfaces with some of the combinations of $A$ and $D$ used in Fig.~\ref{figCompareBidan}b. The tissue interface is shown at days 7, 14, 21 and colored according to cell density. Left column: experimental bioscaffold tissue growth showing the extent of new tissue at days 7 (red), 14 (orange), and 21 (yellow) (reproduced with permission from \cite{Bidan2012,Bidan2013}).}
\label{figCompareBidan2}
\end{figure*}
Clearly, the depletion rate of active cells $A$ strongly influences tissue deposition slowdown (Fig.~\ref{figCompareBidan}b). While diffusivity $D$ has only a weak influence on $A_\text{T}(t)$, it drives important qualitative changes in the shape of the tissue interface and in the distribution of cells on the interface, as in the simulations presented in Section~\ref{Sect4a}. Figure~\ref{figCompareBidan2} compares the experimental evolution of the tissue surface in the semi-circular and cross-shaped pores with that predicted by the cell-based model with some of the combinations of $A$ and $D$ used in Fig.~\ref{figCompareBidan}b. While both combinations $A=0.1/\da$, $D=0.0001\,\mm^2/\da$ (blue) and $A=0.1/\da$, $D=0.01\,\mm^2/\da$ (cyan) give similar growth curves $A_\text{T}(t)$ in Fig.~\ref{figCompareBidan}b, the evolution of the tissue interface obtained with $D=0.0001\,\mm^2/\da$ is much closer to the experimental tissue surface in Fig.~\ref{figCompareBidan2}. The values $A=0.1/\da$, $D=0.0001\,\mm^2/\da$ are shown in Appendix~\ref{appx:errorcalc} to minimise an error function that combines discrepancies both in tissue produced and in shape of the interface.

\section{Discussion}
\label{Sect5}
The explicit consideration of the cellular origin of new tissue growth enables us to model a systematic influence of local curvature on cell density and growth rate. During the evolution of bone tissue \emph{in vivo}, this influence represents the inevitable geometrical pull of the local expansion or contraction of curved bone surfaces. This effect is important to assess in order to understand the emergence of various formation patterns seen in bone histology in anthropological studies~\cite{Maggiano2011,Maggiano2016} and to correctly quantify the influence of other processes on tissue growth. During osteonal infilling for example, surface area shrinks to about 20\% of its initial extent, yet the density of active osteoblasts depositing new bone decreases. Area shrinkage is strongly overpowered by depletion pathways from the pool of active osteoblasts~\cite{Parfitt1994,Pazzaglia2011,Buenzli2014a}. Mathematical models of multistage osteoblast development have modelled these different contributions in previous works~\cite{Polig1990,Buenzli2014a}, but they were restricted to perfectly cylindrical infilling cavities.

Here, we show that such cell-based models can explain both smoothing of irregular initial substrates and tissue deposition slowdown. The co-evolution of tissue interface and cell density exhibits rich behaviours depending on the strength of cell diffusion along the interface~\cite{Crank1975} and on the depletion rate of active cells. This is due in part because cells diffusing on stretching domains may or may not colonise them depending on the ratio of diffusivity and domain growth~\cite{Simpson2006}. Here, cell density inhomogeneities induced by stretch additionally drive the evolution of domain stretch. Mathematically, our equations form a class of hyperbolic curvature flow~\cite{Sethian1999} rather than mean-curvature flow~\cite{Grayson1987, Crandall1983}. As a result, cusps may emerge in finite-time in the zero-diffusion limit. Curvature and cell density behave similarly to the conjugate variables of a harmonic oscillator (such as position and velocity). Shock waves and inertial effects leading to oscillatory interface motion occur for low enough diffusive damping (Figs~\ref{figCosine}--\ref{figVariousShapes}). These shocks and oscillatory motions involve strong inhomogeneities in cell density (Fig.~\ref{figCosine}b). They could represent some patterns of stepwise lamellar sheet bone formation at a large scale~\cite{Maggiano2011}, though it is also possible that these lamellar sheets are formed discontinuously in time. At a smaller scale, it is likely that cell density does not develop long-lasting inhomogeneities in space. Records of the forming bone surface provided by lamellae in cortical osteons~\cite{Pazzaglia2012}, primary bone, and between curved trabecular structures in corticalised bone~\cite{Maggiano2011,Streeter2011,Maggiano2016} display efficient smoothing and the absence of centred cusps in concavities, as in our simulations with intermediate cell diffusivity.

The density of osteoblasts on active bone surfaces is not often measured~\cite{Marotti1976}, making it difficult to disentangle the contributions of cell density and cell vigor to the normal velocity of the interface, called matrix apposition rate in Biology~\cite{Parfitt1983,Buenzli2014a}. Osteoblast density is influenced by the transition to non-synthesising, tissue-embedded cells called osteocytes. The sink term that describes the depletion of the pool of active cells at rate $A$ in Eq.~\eqref{rho_t3} also models the transition to such tissue-embedded cells~\cite{Buenzli2015}. In Buenzli~\cite{Buenzli2015}, it is shown that osteocyte density does not depend explicitly on osteoblast density, only on the ratio of the rate of osteoblast burial and secretory rate. One should therefore not regard a homogenous distribution of embedded osteocytes as a sign that osteoblast density was homogeneous. In fact, some degree of inhomogeneous osteoblast density is likely. Osteoblasts are believed not to move significantly with respect to the bone surface as they have several cellular projections linking with osteocytes through the bone tissue matrix~\cite{Parfitt1994,Pazzaglia2014}.

Tissue growth in bioscaffolds is less polarised than bone apposition \emph{in vivo}. Cells proliferate and may produce extracellular matrix in random directions to create new tissue. However, fibronectin labelling used recently by Bidan~\etal\ suggests that bulk tissue does not swell or compress during its maturation~\cite{Bidan2016}. Deep fibronectin labels are stationary and the density of embedded cells is homogeneous, showing that new tissue production is concentrated near the tissue surface, possibly as a result of increased tissue tension there~\cite{Bidan2016} that could promote cell proliferation~\cite{Nelson2005}. The geometrical influence of curvature captured by our equations also holds in this situation. New cellular and extracellular tissue produced near concave portions of the surface will accelerate the velocity of the local interface in proportion to curvature (Fig.~\ref{figCell}a). Our numerical simulations show that this influence leads to a very good match with experimental tissue growth patterns and slowing rates (Figs~\ref{figCompareBidan}, \ref{figCompareBidan2}). In these simulations, the increased crowding of tissue produced in concavities leads to smoothing, while the depletion of active cells leads to tissue deposition slowdown. Depleting the pool of active cells corresponds to the hypothesis that cells slow down, and eventually stop, the production of new tissue as they find themselves deeper within the tissue and mature~\cite{Bidan2012}. It should also be noted here that cell proliferation was assumed in the simulations to be balanced out by the transition to quiescent tissue-embedded cells, with an overall net depletion of active cells as described by the negative first-order reaction rate in Eq.~\eqref{rho_t3}.

Tissue surface tension has been considered to play a role in bioscaffold tissue growth~\cite{Bidan2012,Nelson2005,Rumpler2008,Dunlop2010,Bidan2013,Bidan2013b,Gamsjager2013}. Surface tension accounts for the relaxation of membranes towards minimal surfaces by curvature-controlled flow. In the thermodynamically consistent mechanical model of tissue growth of Ref.~\cite{Gamsjager2013}, surface tension was added to explain that new tissue could not be produced on convex substrates unless the chemical growth force dominated the surface stress, which works in the opposite direction at convexities.  While surface tension due to the dense actin network near the tissue surface may play a mechanical role in the tissue's growth rate at concavities~\cite{Bidan2012,Gamsjager2013}, we did not consider this effect here, and focused on how new tissue volume is created and fills available space. Our approach is similar to the (compressive) stress-dependent eigenstrain tissue growth model of Refs~\cite{Dunlop2010,Gamsjager2013} except that we directly consider the volumetric crowding of tissue rather than the mechanics-induced movement of tissue created by its volumetric growth rate. Doing so enables us to exhibit an explicit dependence of the tissue interface motion upon local curvature (without surface tension). This dependence occurs via the normal acceleration of the tissue interface and leads to oscillatory behaviour at low damping. While the model developed by Dunlop and co-workers in Refs~\cite{Dunlop2010,Gamsjager2013} has been applied to circular pore shapes with rotation symmetric solutions only, it is also possible that the thermodynamic dissipation assumed in their model would disallow oscillatory motions. 

Complex growth patterns also occur in problems of interfacial thermodynamics and in diffusion-limited aggregation. In these systems, growth is mostly determined by diffusive fluxes external to the growing substrate and by surface tension (e.g., via the Gibbs--Thomson relation)~\cite{Langer1980}. External nutrient fluxes, surface tension, and mechanical loading~\cite{Guyot2014,Guyot2015} may of course add further dependences of tissue growth on curvature, in particular via cell vigor. The curvature influence on density exhibited by our model must be singled out to assess the true impact of these effects.

In summary, the shrinking or expanding available space near concavities or convexities of growing tissues provides an unavoidable geometric influence in a number of situations in which tissue production occurs near the interface, such as in tumour growth, wound healing, bone formation, and bioscaffold tissue growth. We showed that this influence is captured as a curvature-dependent acceleration of tissue growth. In bioscaffold tissue growth, contractile tension may further help even out cell densities and extracellular matrix, enhancing the smoothing dynamics. During bone formation \emph{in vivo}, cellular tissue tension is likely to play a more minor role. Bone matrix quickly mineralises and osteoblasts have been shown not to proliferate after becoming active~\cite{Parfitt1983,Parfitt1994}.

% However, tissue surfaces are not free membranes---their evolution is linked to the creation of new tissue volume beneath them. Contractile cell tension has been associated with proliferation in confluent cell sheets~\cite{Nelson2005}, which could explain tissue volume creation, but this effect is small and it may not dominate growth to confluence.
% In Ref.~\cite{Nelson2005}, stress concentrates in the corners of bioscaffolds and at edges of the cell sheet. However in Bidan et al., no new tissue/proliferation occurs deep within tissue near edges of the bioscaffold.

Finally, we suggest that local changes in surface area during the evolution of the interface play a wider role than physically concentring or spreading local cell densities. We hypothesise that dynamic surface area changes may be a mechanism by which cells on a substrate can perceive geometrical features that are much larger than the cells. Cells may not sense these geometrical features directly, but they may sense them dynamically when the interface contracts or expands, either because of changes in cell--cell contact pressure with neighbouring cells, or because of stretching of focal adhesion sites.

\paragraph{Authors' contributions} Conceived and designed the study: AA, PRB. Performed the numerical simulations: AA. Analysed the data: AA, PRB. Drafted the manuscript: AA. All authors edited the manuscript and gave final approval for publication.

\paragraph{Competing interests} We have no competing interests.

\paragraph{Funding} PRB gratefully acknowledges the Australian Research Council for Discovery Early Career Researcher Fellowship (project number DE130101191).

\paragraph{Acknowledgments}
We thank John~Dunlop, J\'er\^ome Droniou, Julie Clutterbuck, Isabel Maggiano and Corey Maggiano for fruitful discussions. We are thankful to the anynomous reviewers whose comments helped improve and clarify the manuscript.

%\clearpage \newpage

\begin{appendices}% using package appendix for more flexibility.

\section{Governing equations}\label{appx:gov-eq}

\subsection{Evolution of the local surface stretch}
\label{appx:stretch}
The local surface stretch in the normal direction $\dperp{}{t}\deltaup\ell$ can be calculated using an orthogonal parameterisation $\b{\Gamma}(u,t)$ of $S(t)$, defined such that ${\b \Gamma}_t  =  v {\b n}$ and ${\b \Gamma}_t \cdot {\b \Gamma}_u=0$ at all times (see Fig.~\ref{figCell}b). Because trajectories $t\mapsto \b\Gamma(u,t)$ (at $u$ constant) are normal to $S(t)$ at all times, $\dperp{}{t} \equiv \left.\pd{}{t}\right)_{u=\text{const}}$. With $\deltaup \ell=G \der u$, where $G=|{\b \Gamma}_u|$ is the metric associated with $\b{\Gamma}(u,t)$, one thus has $ \dperp{}{t}\deltaup \ell \ = G_t \der u$. Differentiating the definition of $G$ and using the unit tangent vector $\b\tau = \b\Gamma_u/|\b\Gamma_u|$ gives $G_t = {\b \tau} \cdot {\b \Gamma}_{ut} = {\b \tau} \cdot \left({\b \Gamma}_{t} \right)_u = {\b \tau} \cdot \left(v {\b n} \right)_u = v \b\tau\cdot \b n_u$ where the last equality uses the fact that $\b\tau\cdot\b n=0$. By definition of the signed curvature, $\kappa = - \b\tau_\ell \cdot \b n = \b \tau \cdot \b n_\ell$ where $\pd{}{\ell}=\frac{1}{G}\pd{}{u}$~\cite{Dineen2014}. Therefore, $G_t = v G \kappa$ and
\begin{align}
	\frac{\dperp{}{t}\deltaup\ell}{\deltaup \ell}
	= \frac{G_t}{G}  
	= v \kappa    \label{eqn8bis}.
\end{align}

\subsection{Conservative form of the governing equations}
\label{appx:conservative}
Equation~\eqref{gamma_t} only specifies the normal component of $\b\gamma_t$. The tangential component $\b\tau\cdot\b\gamma_t$ can be chosen arbitrarily without modifying the evolution of $S(t)$. Once this choice is made, $\b\gamma$ satisfies:
\begin{align}\label{gamma_t_2}
    \b\gamma_t = v\b n +  (\b\tau\cdot\b\gamma_t)\b \tau
\end{align}
where $\b\tau = \b\gamma_s/|\b\gamma_s|$. For choices of $\b\tau\cdot\b\gamma_t$ corresponding to thickness functions in Cartesian and polar coordinates, a conservative equation can be derived by differentiating Eq.~\eqref{gamma_t_2} with respect to $s$ (see Sections~\ref{appx:cartesian}--\ref{appx:polar} below).

To obtain the general conservative form of the evolution equation of cell density, it is necessary to consider the density of cells projected onto the $s$ coordinate, $\eta(s,t)=\rho(s,t)g(s,t)$ (see Section~\ref{Sect2}). We first derive the evolution equation of the local stretch $g$. Differentiating the definition of $g=|\b\gamma_s|$ with respect to $t$ as in Sect.~\ref{appx:stretch} gives, using Eq.~\eqref{gamma_t_2} and the fact that $\b\tau\cdot\b n = 0$, $\b\tau\cdot \b\tau_s = 0$, and $\kappa = \b\tau\cdot\b n_s/g$:
\begin{align}\label{g_t}
    g_t = \b\tau \cdot (\b\gamma_t)_s = v \b\tau\cdot \b n_s + (\b\tau\cdot\b\gamma_t)_s = g v \kappa + (\b\tau\cdot\b\gamma_t)_s.
\end{align}
Equation~\eqref{g_t} generalises the second equality in Eq.~\eqref{eqn8bis} to non-orthogonal parameterisations. Now differentiating $\eta=\rho g$ with respect to $t$, and using Eqs~\eqref{rho_t3} and~\eqref{g_t} gives Eq.~\eqref{eta_t}:
\begin{align}
	\eta_t + \left(-\frac{\eta}{g}({\b \gamma}_t \cdot \b{\tau})  
	- \frac{D}{g} \left(\frac{\eta}{g} \right)_s\right)_s = - A\eta  . \label{eqn13bis}
\end{align}
This equation expresses the balance of cells on the interface element lying between the coordinates $s$ and $s+\der s$. It is of the form $\eta_t + \big(f[\eta,\b\gamma]\big)_s = - A\eta$ with total flux
\begin{align}
	f[\eta,\b\gamma] = -\frac{\eta}{g}({\b \gamma}_t \cdot \b{\tau})  
	- \frac{D}{g} \left(\frac{\eta}{g} \right)_s \label{eqn14}.
\end{align}
The first term in the flux represents the advection of cells with respect to the $s$ coordinate. Curvature-induced changes in cell density are partly included in this term, and partly included in the evolution of the local stretch $g$ in Eq.~\eqref{g_t}, which must be used to reconstruct the physical cell density $\rho=\eta/g$. For an orthogonal parameterisation of $S(t)$, the first term in the flux is absent, in which case all the curvature-induced changes in $\rho$ come from the evolution of the local stretch $G$ in Eq.~\eqref{eqn8bis}. The second term in the flux corresponds to the diffusion of cells along the interface. The factors $g$ account for the fact that this diffusion is measured for the projected cell density along the $s$ coordinate.

\subsection{Cartesian coordinates}\label{appx:cartesian}
Parameterising $S(t)$ by a height function $y=h(x,t)$ in Cartesian coordinates corresponds to taking $s=x$ and $\b\gamma(x,t)=(x,h(x,t))$. In this case, $\b\tau\cdot\b\gamma_t = v h_x$ and Eqs~\eqref{gamma_t} and~\eqref{rho_t3} become:
\begin{align}
h_t&=v \sqrt{1+h_x^2}\label{h_t_appx}\\
\rho_t &= - \rho v \kappa - \rho_x v \cos\alpha + D
\left(\frac{\rho_{xx}}{g^2} - \rho_x \kappa \cos\alpha \right)
-A\rho   \label{rho_t_appx}
\end{align}
where $v = \kform \rho$, $g=\sqrt{1+h_x^2}$, $\cos\alpha=\b n\cdot \b{\hat x}=-h_x/g$, and $\kappa=-h_{xx}/g^3$. To write this system of coupled equations in conservative form, we define $\sigma = h_x$ and $\eta=\rho g$, so that $\sigma_t = h_{xt} = (h_t)_x = (\kform \eta)_x$. With Eq.~\eqref{eqn13bis} rewritten with these definitions, one obtains the system of equations
\begin{align}
    &h_t = \kform \eta \label{h_t_cons}
    \\& \sigma_t + \left[-\kform \eta \right]_x = 0 \label{sigma_t_cons}
    \\& \eta_t + \left[ - \frac{\kform \sigma \eta^2}{1+ \sigma^2}  + D\left(\frac{\sigma\sigma_x\eta}{(1+\sigma^2)^2}  -  \frac{\eta_x}{1+\sigma^2}\right)\right]_x = - A\eta \label{eta_t_cons}
\end{align} 
% eta_t = - [eta cos(alpha) (v + D kappa) - D eta_x/g^2 ]_x
% with 
%       g=sqrt(1+sigma^2)
%       cos(alpha) = -sigma/g
%       kappa = curvature = -h_xx/(1+h_x^2)^(3/2) = - sigma_x/g^3
Note that Eq.~\eqref{h_t_cons} is decoupled from Eqs~\eqref{sigma_t_cons}--\eqref{eta_t_cons}.

\subsection{Polar coordinates}\label{appx:polar}
Parameterising $S(t)$ by the radius function $r=R(\theta,t)$ in polar coordinates corresponds to taking $s=\theta$ and $\b\gamma(\theta, t) = R(\theta,t) \big(\cos\theta, \sin\theta\big)$. In this case, $\b\tau\cdot\b\gamma_t=-v R_\theta/R$ and Eqs~\eqref{gamma_t} and~\eqref{rho_t3} become:
\begin{align}\label{R_t_appx}
R_t=&-v\sqrt{1+ \left(R_{\theta}/R \right)^2} \\
\rho_t= &-\rho v \kappa - \frac{\rho_\theta}{R} v \cos\alpha + D\left(\frac{\rho_{\theta\theta}}{g^2} - \frac{\rho_\theta}{R}\!\left[\tfrac{2}{g}\!-\!\kappa\right]\!\cos\alpha\right) - A\rho\label{rho_t_appx_polar}
\end{align}
where $v=\kform \rho$, $g=R\sqrt{1+(R_\theta/R)^2}$, $\cos\alpha=\b n\cdot \b{\hat\theta}=R_\theta/g$, and $\kappa=(R^2 - R R_{\theta\theta}+2R_\theta^2)/g^3$. Note that $(2/\!g-\kappa)\cos(\alpha)/R = R_\theta(R+R_{\theta\theta})/g^4$. To write this system of coupled equations in conservative form, we define $\sigma = R_\theta$ and $\eta = \rho g$, so that $\sigma_t = R_{\theta t} = (R_t)_\theta = (-\kform \eta/R)_\theta$. With Eq.~\eqref{eqn13bis} rewritten with these definitions, one obtains the system of equations
\begin{align}
    &R_t = - \frac{\kform \eta}{R} \label{R_t_cons}
    \\& \sigma_t + \left[\frac{\kform \eta}{R} \right]_\theta = 0 \label{sigma_t_cons_polar}
    \\& \eta_t + \left[ \frac{\kform \sigma \eta^2}{R(R^2+ \sigma^2)}  + D\left(\frac{\sigma(R+\sigma_\theta)\eta}{(R^2+\sigma^2)^2}  -  \frac{\eta_\theta}{R^2+\sigma^2}\right)\right]_\theta = - A\eta \label{eta_t_cons_polar}
\end{align} 
% eta_t = [-kform eta^2 (R_theta/R)/g^2 - D eta R_theta(R+R_{theta,theta})/g^4 + D eta_theta /g^2]_theta - A eta
% with 
%       g=R sqrt(1+(sigma/R)^2)
In contrast to the Cartesian case, Eq.~\eqref{R_t_cons} is not decoupled from Eqs~\eqref{sigma_t_cons_polar}--\eqref{eta_t_cons_polar}.

\section{Numerical discretisation}\label{appx:num}
Some aspects of the numerical schemes presented in Section~\ref{Sect2} are detailed here. %The reader is referred to the computer code for full implementation details.

At high diffusivity $D$, we used a semi-implicit finite difference scheme on Eqs~\eqref{h_t_appx}--\eqref{rho_t_appx} (Cartesian coordinates) or Eqs~\eqref{R_t_appx}--\eqref{rho_t_appx_polar} (polar coordinates). Upwinding for all first-order derivatives was based on the sign of $h_x$ (Cartesian) or $R_\theta$ (polar). In Cartesian coordinates for example:
\begin{align}
    \pd{f}{x}(x_i) \approx \begin{cases} \frac{1}{\Delta x}\left[ f(x_i) - f(x_{i-1})\right], &\text{if } h(x_{i-1}) > h(x_{i+1}),
    \\ \frac{1}{\Delta x}\left[ f(x_{i+1}) - f(x_{i})\right], & \text{otherwise}.
\end{cases}
\end{align}
We used explicit forward Euler discretisation in time for advective and reaction terms and implicit backward discretisation for diffusive terms to avoid restrictive Courant--Friedrichs--Lewy (CFL) conditions at high diffusivities~\cite{LeVeque2004}.

At low diffusivity $D$, we used a semi-discrete Kurganov--Tadmor (KT) finite volume method with fully explicit forward Euler time discretisation on Eqs~\eqref{h_t_cons}--\eqref{eta_t_cons} (Cartesian coordinates), or Eqs~\eqref{R_t_cons}--\eqref{eta_t_cons_polar} (polar coordinates). These systems of equations were recast in the form
\begin{align}
    \b u_t + \left[ \b g(\b u) \right]_x = \left[ \b Q(\b u, \b u_x) \right]_x + \b R\label{kt-generic-adv-diff}
\end{align}
for $\b u = (h, \sigma, \eta)$ (Cartesian) or $\b u = (R, \sigma, \eta)$ (polar), where $\b R=(0,0,-A \eta)$. The flux $\b g$ is hyperbolic and contains the part of the total flux that is independent of $\eta_x$. The flux $\b Q$ is parabolic and contains the part that depends on $\eta_x$. The semi-discrete KT form of Eq.~\eqref{kt-generic-adv-diff} at point $x_i$ is
\begin{align}
    \td{}{t}\b u_i(t) = &-\frac{\b H_{i+1/2}(t)-\b H_{i-1/2}(t)}{\Delta t} \notag
\\&+ \frac{\b P_{i+1/2}(t)-\b P_{i-1/2}(t)}{\Delta t} + \b R_i(t),\label{kt-semi-discrete}
\end{align}
where $\b H$ is the Rusanov numerical flux approximating the hyperbolic flux, and $\b P$ is a second order central difference approximation to the parabolic flux (see Eqs~(4.13), (4.14), and (4.4) of Ref.~\cite{Kurganov2000}). The Rusanov fluxes involve left and right values of $\b u_{i\pm 1/2}$ interpolated using a minmod limiter componentwise. The only information on characteristics required in $\b H$ is the maximum absolute value of the eigenvalues of $\b g'(\b u)$ for the Riemann problem at a Riemann fan~\cite{Kurganov2000}, which was determined numerically. Equation~\eqref{kt-semi-discrete} was discretised in time using a simple forward Euler scheme.

 % as done in \cite{Landman2005,Simpson2006}  for a chemotactic and diffusive cell migration problem. This is a high-resolution central scheme, whereas another variant of finite volume scheme is also available known as the upwind or Godunov scheme \cite{LeVeque2004}.

\begin{figure}[!h]
% \captionsetup{justification=centering}

\centerline{
\includegraphics[trim={0 0 30 0},width=1.05\linewidth,clip]{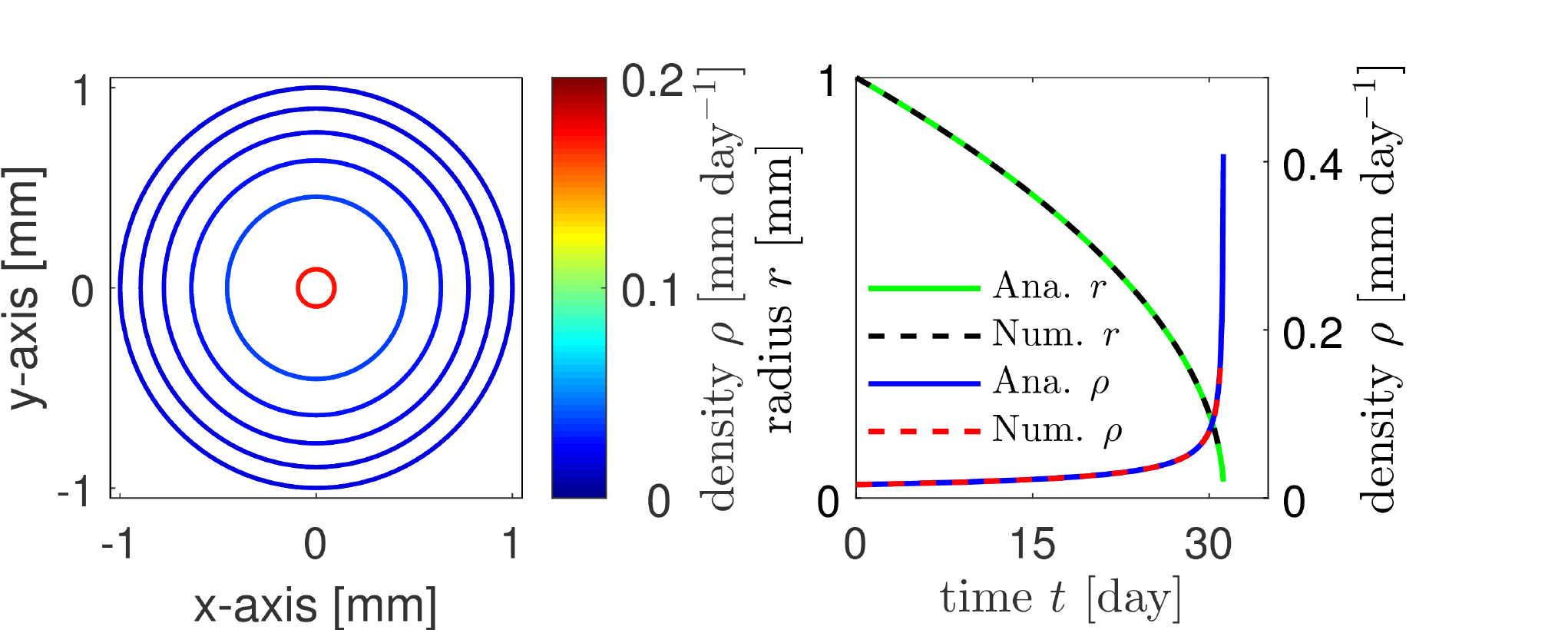}}

\caption{Evolution of circular interface and osteoblast surface density with any cell diffusion value (left). 
There is excellent agreement between the numerical and analytic evolutions of radius and cell density (right). }
\label{fig2}
\end{figure}

Space and time discretisation steps were reduced within constraints imposed by the Courant number~\cite{LeVeque2004} until numerical convergence. Hyperbolic problems of interface propagation are known to give good results even for low order time discretisations. They are more sensitive to spatial discretisation~\cite{Sethian1999}. Note that the Kurganov--Tadmor scheme is a high-resolution central scheme in space. Both numerical schemes were checked against each other for a range of intermediate diffusivities. They were also checked against the analytic solution of the rotation-symmetric infilling circular cavity found from Eqs~\eqref{mcf-circle}--\eqref{calibration1}:
\begin{align}
    R(t) = R_0\sqrt{1-2\frac{v_0}{R_0}\,\frac{1-\exp(-A t)}{A}}.
\end{align}
Figure~\ref{fig2} shows the case where $A=0$, in which $R\to 0$ and $\rho\to\infty$ when $t\to t_c=\frac{1}{2}\frac{R_0}{v_0}\approx 31\,\days$.  Due to symmetry, the solution is independent of cell diffusion and all the numerical solutions are indistinguishable. 

\section{Error function for parameter estimation}\label{appx:errorcalc}
\begin{figure}[!h]
% \captionsetup{justification=centering}

\centerline{
\includegraphics[trim={0 0 0 0}, width=0.8\linewidth,clip]{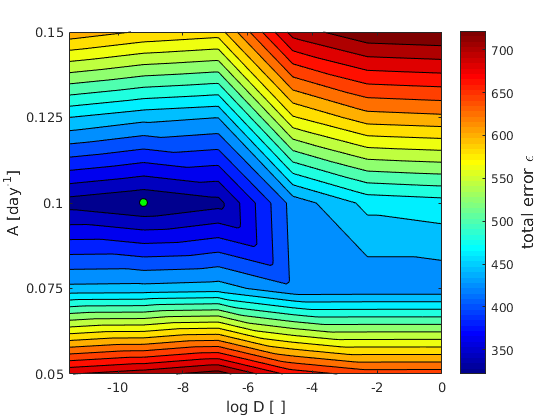}}

\caption{The map of total error $\epsilon$ where $\epsilon = \alpha\ \epsilon_\text{PTA} + \epsilon_\kappa$ and $\alpha \approx 549.6$. The minimum error, marked by the green dot corresponds to $A=0.1$ and $D=\exp(-9,2)\approx 0.0001$.}
\label{fig3}
\end{figure}

Two types of errors were combined to estimate the cell depletion rate $A$ and cell diffusivity $D$ that minimise the discrepancy between numerical simulations and experimental data in Section~3.2. The first error corresponds to the discrepancy in total tissue produced $A_T(p,t)$ summed over the data time points of Fig 5b, and over the four different bioscaffold pore shapes considered, i.e., circular, semi-circular, square, and cross ($p=1,...,4$):
\begin{align}
\epsilon_\text{PTA}(A,D) = \sum_{p=1}^4 \sum_{t} \left |A_T^\text{model}(p,t) - A_T^\text{data}(p,t)\right | .    %\tag{S5}
\end{align} 
Minimising this error ensures a good fit between simulations and data points in Fig.~5b, but this error is only weakly sensitive to values of $D$. In particular, this error does not measure discrepancies in the shape of the interface. To penalise such discrepancies, we considered in addition the least square error of the local curvature $\kappa$ of the last interface available ($t=21\,\days$):
\begin{align}
\epsilon_\kappa(A,D) = \sum_{p=1}^4 \sum_{i} \left |\kappa^\text{model}(p,i) - \kappa^\text{data}(p,i)\right |^2 \Delta \ell_{p,i},     %\tag{S6}
\end{align} 
where $i$ runs over all the discretisation points of the interface, and $\Delta\ell_{p,i}$ is the segmental length between points $i$ and $i+1$. Both error measures are combined into the total error function
\begin{align} \label{total_error}
\epsilon(A,D) = \alpha\ \epsilon_\text{PTA}(A,D) + \epsilon_\kappa(A,D),  %\tag{S7}
\end{align}
where the weight $\alpha\approx 549.6\,\mm^{-3}$ accounts for the difference in unit and order of magnitude of $\epsilon_\text{PTA}$ and $\epsilon_\kappa$, and was set as the ratio between the mean values of $\epsilon_\kappa$ and $\epsilon_\text{PTA}$. A plot of this error surface in the $(D,A)$ parameter space is shown in Figure~\ref{fig3}. The minimum error is obtained for $A\approx 0.1$ and $D\approx \exp(-9.2)\approx 0.0001$. We note here that an objective error function to penalise discrepancies in interface shape is difficult to define. We chose the least square error of local curvature rather than the least square error of interface height or radius because the latter was not very sensitive to $D$ and similar to $\epsilon_\text{PTA}$. Other variations are possible, and may lead to slightly different optimal values of $A$ and $D$.

\end{appendices}

\bibliographystyle{vancouver}
\bibliography{reference}

\end{document}